\documentclass[12pt]{article}
\usepackage[latin1]{inputenc}
\usepackage{amsmath,amssymb,amsfonts}
\usepackage{undertilde}
\usepackage{fullpage}
\usepackage{graphicx}
\newtheorem{theorem}{Theorem}[section]
\newtheorem{lemma}[theorem]{Lemma}
\newtheorem{proposition}[theorem]{Proposition}

\begin{document}
\title{The exp-$G$ family of probability distributions}
\author{Wagner Barreto-Souza${}^{a}$ and Alexandre B. Simas${}^{b}$\\
${}^a$ Departamento de Estat\'istica, Universidade de S\~ao Paulo,\\ 
Rua do Mat\~ao, 1010,
S\~ao Paulo/SP, 05508-090, Brazil\\
e-mail: {wagnerbs85\@ hotmail.com}\\
${}^{b}$ Associa\c{c}\~ao Instituto Nacional de Matem\'atica Pura e Aplicada, IMPA,\\
Estrada D. Castorina, 110, Jd. Bot\^anico,
Rio de Janeiro/RJ, 22460-320, Brazil\\
e-mail: {alesimas\@ impa.br}}
\date{}
\maketitle

\begin{abstract}
In this paper we introduce a new method to add a parameter to a family of distributions. The additional parameter is completely studied and a full description of its behaviour in the distribution is given. We obtain several mathematical properties of the new class of distributions such as Kullback-Leibler divergence, Shannon entropy, moments, order statistics, estimation of the parameters and inference for large sample. Further, we showed that the new distribution have the reference distribution as special case, and that the usual inference procedures also hold in this case. Furthermore,
we applied our method to yield three-parameter extensions of the Weibull and beta distributions. To motivate the use of our class of distributions,
we present a successful application to fatigue life data.\\
\emph{Keywords}: Exp-$G$ distribution; Order statistics; Fisher's information matrix; Exp-Weibull distribution; Exp-beta distribution.
\end{abstract}

\section{Introduction}
The present work is an enhanced and extended version of the pioneering manuscript presented at Est\^ancia de S\~ao Pedro, S\~ao Paulo, Brazil, in the 18º SINAPE, 2008, see Barreto-Souza et al. (2008). 

In many practical situations, usual probability distributions do not
provide an adequate fit. By example, if the data are asymmetric, normal 
distribution will not be a good choice. With this, several methods of introducing
a parameter to expand a family of distributions have been studied. 

Marshall and Olkin (1997) introduced a new way to expand probability distributions 
and applied to yield a two-parameter extension of the exponential distribution which
can serve as a competitor to such commonly-used two-parameter distributions as the 
Weibull, gamma and lognormal distributions. Furthermore, this method was used to obtain a three-parameter extension of the Weibull distribution. Moreover,
Mudholkar et al. (1996) introduced a three-parameter distribution alternative to the Weibull distribution, that has the Weibull as limiting distribution. 

Some methods of introducing of parameters to symmetric distributions have been studied in order to add skewness. For instance, Azzalini (1985) introduced 
and studied the well-known skew-normal distribution, which is obtained by adding a shape parameter to the normal distribution. Another symmetric distribution that was extended by adding a skewness parameter was the Student's-$t$ distribution by Jones and Faddy (2003). Finally, Ma and Genton (2004) introduced a general class of skew-symmetric distributions, whereas Ferreira and Steel (2006) provides a general perspective on the introduction of skewness into symmetric distributions.  

Recently, Jones (2004) introduced a class of distributions that adds two parameters to a reference distribution. Further, Jones and Pewsey (2009) advanced a
four-parameter family has both symmetric and skewed members and allows for tail weights that are both heavier and lighter than those of the generating
distribution.
 
 In this article, we introduce a new method to add a parameter to some reference distribution. The resulting distribution exhibits the remarkable reciprocal property. We study this parameter in detail, and we give a full description of its behaviour in the distribution. The augmented distribution has several connections with the reference distribution, for instance, the Kullback-Leibler divergence of the augmented distribution with respect to the original distribution is finite and only depends on the new parameter. Several others properties in this direction are also given. The inferential aspects of this distribution are studied in details, and two special cases are discussed, and a successful empirical application shows the flexibility the new distribution, and also motivates its usage.
 
A special attention must be given to the fact that it is not straightforward that this new distribution contains the reference distribution as special case. We show that this is the case if we enlarge the parameter space, and also, that this enlargement is good, in the sense that, all the standard inferential procedures work if this new value in the parameter space is considered to be the true value of the parameter.
 
 The remaining of the article unfolds as follows: in Section \ref{sec2} the new class of distributions is introduced, several properties are given, the new parameter is completely characterized, and the inferential aspects are discussed. Sections \ref{sec3} and \ref{sec4} deals with two special cases: the exp-Weibull and exp-beta distributions, respectively. In Section \ref{sec5} an empirical application shows the usefulness of this distribution. Finally, Section \ref{sec6} ends the article with some concluding remarks. The Appendix contains the proofs of the results presented in the article.

\section{The new class of distributions}\label{sec2}

The cdf of a random variable with truncated exponential distribution in the interval $[0,1]$ with parameter $\lambda$ is given by
\begin{eqnarray}\label{trunc}
F_\lambda^{*}(x)=\frac{1-e^{-\lambda x}}{1-e^{-\lambda}},  
\end{eqnarray}
where $\lambda>0$ and $x\in[0,1]$. We now observe that $F_\lambda^{*}(\cdot)$ is a cdf for $\lambda\in\mathbb{R}\setminus\{0\}$, and that
$$\lim_{\lambda\to 0} F_\lambda^{*}(x) = x,\qquad x\in [0,1].$$
Therefore, we extend the parameter space of the distribution above for the entire line:
$$F_\lambda(x) = \left\{\begin{array}{cc}
F_\lambda^{*}(x),&\hbox{if~~}\lambda\neq 0,\\
x,&\hbox{if~~~}\lambda=0.\end{array}\right.$$

We now define the new class as follows. Let $G(x;\theta)$ be the cdf of a continuous or discrete random variable with $\theta$ being the parameters related to $G$, then the class of distributions exp-$G$, indexed by $\lambda$, is defined by
\begin{eqnarray}\label{expg}
F_\lambda^G(x)=F_\lambda(G(x;\theta)).
\end{eqnarray}

From now on, we will denote a random variable $X$ with cdf (\ref{expg}) by $X\sim\exp\mbox{-}G(\Theta)$, where $\Theta=(\lambda,\theta)^T$. 

If $G(x;\theta)$ is a cdf of a continuous random variable, then the exp-$G$ distribution is absolutely continuous for every $\lambda\neq 0$, and its probability density function (pdf), which is the derivative of the cdf (\ref{expg}) with respect to $x$, is given by
\begin{eqnarray}\label{continuous}
 f(x)\equiv f_\lambda(x)=\frac{\lambda}{1-e^{-\lambda}}g(x;\theta)\exp\{-\lambda G(x;\theta)\},
\end{eqnarray} 
where $g(\cdot;\theta)$ is the pdf associated to the cdf $G(\cdot;\theta)$. Let $G(x;\theta)$ be a cdf of a discrete random variable taking values on the set $\{x_1,x_2,\ldots\}$, where $x_1<x_2<\cdots$, then the corresponding exp-$G$ distribution is also discrete, takes values on the same set for every $\lambda\neq 0$, and its probability function is given by
\begin{eqnarray}\label{discrete}
P_\lambda(x_i)=\frac{\exp\{-\lambda G(x_{i-1};\theta)\}-\exp\{-\lambda G(x_{i};\theta)\}}{1-e^{-\lambda}},
\end{eqnarray}
where $G(x_0)=0$.

If $G(x;\theta)$ is a absolutely continuous cdf, then its hazard function is given by
\begin{eqnarray}\label{haz}
h(x;\theta)\equiv h_\lambda(x;\theta)=\frac{\lambda g(x;\theta)}{1-\exp\{-\lambda S(x;\theta)\}},
\end{eqnarray}
where $S(x;\theta)=1-G(x;\theta)$ is the survival function of a random variable with cdf $G(\cdot,\theta)$.

We now state several results regarding the relation between the exp-$G$ and $G$ distributions, where proofs can be found in the Appendix.

\begin{proposition}\label{prop1}
Let $X$ and $X_\lambda$ have $G$ distribution and exp-$G$ distribution with parameter $\lambda$, respectively. Let also $\mu$ be the law of $X$, and $\mu_\lambda$ be the law of $X_\lambda$. Then,
\begin{itemize}
\item[i)] $X$ and $X_\lambda$ have the same support for all $\lambda\neq 0$;
\item[ii)] if $X$ is continuous, singular or discrete, then $X_\lambda$ is continuous, singular or discrete, respectively, for all $\lambda\neq 0$;
\item[iii)] $\mu <\!< \mu_\lambda$, that is, $\mu_\lambda$ is absolutely continuous with respect to $\mu$. Moreover, the Radon-Nikodym derivative of $\mu_\lambda$ with respect to $\mu$ is, almost surely,
$$\frac{d\mu_\lambda}{d\mu}(x) = \frac{1}{1-e^{-\lambda}}\lim_{\substack{\epsilon\to 0\\\epsilon>0}} \frac{\exp\{-\lambda G(x-\epsilon)\}-\exp\{-\lambda G(x)\}}{G(x)-G(x-\epsilon)};$$
\item[iv)] if $X$ is continuous, the relative entropy (Kullback-Leibler divergence) between $X$ and $X_\lambda$ is
$$D_{KL}(\mu\mid\mid\mu_\lambda) = -\int\log\frac{d\mu_\lambda}{d\mu} d\mu = 1-\frac{\lambda}{e^\lambda-1} - \log\left(\frac{\lambda}{1-e^{-\lambda}}\right);$$
\item[v)] if $E(|X|^r)<\infty$, then $E(|X_\lambda|^r)<\infty$, and, moreover, if $\lambda>0$
$$E(|X|^r) \geq \frac{\lambda}{1-e^{-\lambda}}E(|X_\lambda|^r),$$
and if $\lambda<0$
$$\frac{\lambda}{1-e^{-\lambda}} E(|X|^r) \leq E(|X_\lambda|^r).$$
\end{itemize}
\end{proposition}

We now give a characterization for our class of distributions through Shannon entropy. Such entropy were introduced by Shannon (1948) and, for a random variable $X$ with density $f(\cdot)$, with respect to a $\sigma$-finite measure $\mu$, usually the Lebesgue or counting measure, is given by
\begin{eqnarray}\label{sh}
\mathbb{H}_{S}(f)=-\int_{\mathbb{R}}f(x)\log f(x)d\mu.
\end{eqnarray}
 Jaynes (1957) introduced one of the most powerful techniques employed in the field of probability and statistics called maximum entropy method. This method is closely related to the Shannon entropy and considers a class of density functions
\begin{eqnarray}\label{class}
\mathbb{F}=\{f(x): E_f\{T_i(X)\}=\alpha_i, \quad i=0,\ldots,m\}
\end{eqnarray}
where $T_i(X)$, $i=1,\ldots,m$, are absolutely integrable functions with respect to $fd\mu$, and $T_0(X)=\alpha_0=1$. In the continuous case, the maximum entropy principle suggests to derive the unknown density function of the random variable $X$ by the model that maximizes the Shannon entropy in (\ref{sh}), subject to the information constraints defined in the class $\mathbb{F}$. 

The maximum entropy distribution is the density of the class $\mathbb{F}$, denoted by $f^{ME}$, which is obtained as the solution of the optimization problem 
\begin{eqnarray*}
f^{ME}=\mbox{arg} \max_{f\in\mathbb{F}} \mathbb{H}_{S}(f).
\end{eqnarray*}
Jaynes (1957), in the page 623, states that the maximum entropy distribution $f^{ME}$, obtained by the constrained maximization problem described above, ``is the only unbiased assignment we can make; to use any other would amount to arbitrary assumption of information which by hypothesis we do not have.'' It is the distribution which should not incorporate additional exterior information other than which is specified by the constraints.

In order to obtain a maximum entropy characterization for our class of distributions, we now derive suitable constraints. For this, the next result plays a important role. We will assume in the Propositions \ref{prop2} and \ref{prop3} that the reference measure, $\mu$, is the Lebesgue measure, and that all the random variables involved are continuous.\\

\begin{proposition}\label{prop2}
Let $G$ be the distribution of a continuous random variable, with pdf, $g(\cdot)$, and let $X$ be a random variable with pdf $f(\cdot)$ given by \ref{continuous}. Then, we have that
\begin{eqnarray*}
C1)&& \,\,\,E\{\log g(X;\theta)\}=E\{\log g(G^{-1}(U;\theta))\}=\frac{\lambda}{1-e^{-\lambda}}\int_0^1 \log g(G^{-1}(u;\theta))e^{-\lambda u}du,\\
C2)&& \,\,\, E\{G(X;\theta)\}=\frac{1}{\lambda}-\frac{1}{e^\lambda-1}
\end{eqnarray*} 
and the Shannon entropy of $f(\cdot)$ is given by
\begin{eqnarray}\label{entropy}
\mathbb{H}_{S}(f)=1-\frac{\lambda }{e^\lambda-1}-\log\left(\frac{\lambda}{1-e^{-\lambda}}\right)-E\{\log g(G^{-1}(U;\theta))\},
\end{eqnarray}
where $U$ follows truncated exponential distribution with parameter $\lambda$ and cdf given by (\ref{trunc}).
\end{proposition}

The next proposition shows that the class exp-$G$ of distributions has maximum entropy in the class of all probability distributions specified by the constraints stated therein.\\

\begin{proposition}\label{prop3} The pdf $f(\cdot)$ of a random variable $X$, given by (\ref{continuous}), is the unique solution of the
optimization problem
\begin{eqnarray*}
f =\mbox{arg} \max_{h\in\mathbb{F}} \mathbb{H}_{S}(h),
\end{eqnarray*}
under the constraints C1 and C2 presented in Proposition \ref{prop2}.
\end{proposition}

\subsection{$\lambda$ as a concentration parameter}

We provide two asymptotic results of this class, by making the parameter $\lambda$ tend to $\pm\infty$. This results will allow us to give an interpretation for this parameter. Since $F_\lambda(x)\to x$ as $\lambda\to 0$, we have, trivially, that if $X_\lambda^G\sim{\rm exp}-G$ and $X^G\sim G$, then
$$X_\lambda^G \stackrel{d}{\longrightarrow} X^G,$$
as $\lambda\to 0$, where $\stackrel{d}{\longrightarrow}$ stands for convergence in distribution.

Therefore, the definition of the family exp-$G$ by using (\ref{expg}) with $\lambda\in \mathbb{R}$ is good. This fact plays an important role in our paper because this makes the family exp-$G$ contain $G$ as particular case. The following result is very important since regular distributions in Statistics enjoy many desirable properties.\\

\begin{proposition}
If $G$ is a parametric regular probability distribution, with parametric space $\Theta$, then so is the exp-$G$ distribution, with respect to the parametric space $\mathbb{R}\times\Theta$.
\end{proposition}
{\it Proof.}\\
The proof follows from a simple verification of the conditions given in Lehmann and Casella (2003).\\
${}$ \hfill $\square$

The distribution may present very different behaviour for large absolute values of $\lambda$, thus showing that this is a rich class of distributions.

Going further on the discussion of what happens when the absolute value of $\lambda$ is large. We begin by noting that $F_\lambda^G$ will tend to one if $\lambda$ tends to infinity, whenever $x$ is such that $G(x)>0$, and will be zero otherwise. Therefore, if $X_\lambda$ follows a exp-$G$ distribution, where $G$ is any cdf, then
$$X_\lambda \stackrel{v}{\longrightarrow} \delta_{a},$$
as $\lambda\to\infty$, where $a=\inf\{x; G(x)>0\}$, `$\stackrel{v}{\longrightarrow}$' stands for vague convergence, and $\delta_a$ is the Dirac's measure concentrated on $a$, that is, $\delta_a (\{a\}) = 1$. Note that we needed to consider the vague convergence instead of convergence in distribution to allow $a=-\infty$. If $a=-\infty$, then
$$X_\lambda \stackrel{v}{\longrightarrow} 1,$$
where $1$ is the function identically equal to one, which is not a probability measure. However, we may interpret this case as a ``probability measure'' concentrated at $-\infty$, that is, if a random variable would follow $1$, then $\mbox{pr}(X\leq x) = 1$ for all $x\in \mathbb{R}$. 

We now obtain the asymptotic behaviour of $\lambda\to-\infty$. For this case, a simple calculus argument allows us to conclude that $F_\lambda^G$ will tend to zero, whenever $x$ is such that $G(x)<1$, and will be 1 otherwise. Therefore,
if $X_\lambda$ follows a exp-$G$ distribution, where $G$ is any cdf, then
$$X_\lambda\stackrel{v}{\longrightarrow} \delta_{b},$$
as $\lambda\to -\infty$, where $b=\sup\{x; G(x)<1\}$. Note that we also needed to use the vague convergence to include the case where $b=\infty$. In this case
$$X_\lambda \stackrel{v}{\longrightarrow} 0,$$
where $0$ is the function identically equal to zero, which, again, is not a probability measure. However, we may, accordingly, interpret this case as a ``probability measure'' concentrated at $\infty$, that is, if a random variable would follow $0$, then $\mbox{pr}(X\leq x) = 0$ for all $x\in \mathbb{R}$. 

We see from this result, that the parameter $\lambda$ can be interpreted as a concentration parameter, because it moves the exp-$G$ distribution to a degenerated distribution in $a$ (if $a$ is finite), when it varies from zero to infinity, and to a degenerated distribution in $b$ (if $b$ is finite) when it varies from 0 to minus infinity. Furthermore, if $a$ equals minus infinity, the distribution moves towards the left side of the axis until the mass escape entirely, when $\lambda$ tends to infinity. Analogously, when $b$ equals infinity, the distribution moves towards the right side of the axis until the mass escape entirely, when $\lambda$ tends to minus infinity.

\subsection{Reciprocal property}

This family of distributions enjoys a very interesting reciprocal property. We begin by introducing some notation, let $X^G\sim G$, and $1/X^G\sim S$, where $G$ is continuous. Therefore, we have that if $X_\lambda^G\sim{\rm exp-}G$, then $1/X_\lambda^G\sim{\rm exp-}S$. To see this, observe that, for $\lambda\neq 0$,
$$\mbox{pr}(1/X_\lambda^G \leq x) = \mbox{pr}(X_\lambda^G\geq 1/x) = \frac{1-\exp\{\lambda(1-G(1/x))\}}{1-\exp\{\lambda\}} = F_{-\lambda}^{*}(S(1/x)).$$
We also would like to remark that the reciprocal of $X_\lambda^G$ has a corresponding exp-$S$ distribution with $-\lambda$, that is, $X_\lambda^G$ has cdf $F_\lambda^G(x)$ and $1/X_\lambda^G$ has cdf $F_{-\lambda}^S(x)$. 

 This means that whenever we study a special case of the exp-$G$ distribution, we may easily study the reciprocal case.
For instance, in this paper we study the exp-Weibull, and from this result,  we also obtain several properties of the exp-Fréchet distribution.

\subsection{Expansions, order statistics and moments}\label{expansions}

We now give an useful expansions for the pdf (\ref{continuous}). With this expansion, we can obtain mathematical properties such as ordinary moments, factorial moments and moment generating function of the exp-$G$ distribution from $G$ distribution. Expanding the term $e^{-\lambda G(x;\theta)}$ in (\ref{continuous}), it follows
\begin{eqnarray}\label{exppdf}
f(x)=\frac{\lambda}{1-e^{-\lambda}}g(x;\theta)\sum_{j=0}^\infty \frac{(-\lambda)^j}{j!}G(x;\theta)^j.
\end{eqnarray}
If $G(\cdot;\theta)$ has not closed-form, suppose 
\begin{eqnarray}\label{expcdf}
G(x;\theta)=\sum_{k=0}^\infty a_k x^{k+c},
\end{eqnarray}
 where
$\{a_k\}_{k=0}^\infty$ is a sequence of real numbers and $c\in \mathbb{R}$. Several distributions do not have closed-form cdf and can be written in the form (\ref{expcdf}), we have, for instance, the normal, gamma and beta distributions.\\

For $n$ positive integer, we have 
\begin{eqnarray}\label{expnint}
\left(\sum_{k=0}^\infty a_k x^k\right)^n=\sum_{k=0}^\infty c_{n,k} x^k,
\end{eqnarray}
where $c_{n,0}=a_0^n$ and  $c_{n,m}=(m a_0)^{-1}\sum_{k=1}^{m}(nk-m+k)a_k c_{n,m-k}$ for 
$m=1,2,\ldots$ (see Gradshteyn and Ryzhik, 2000). Using (\ref{expcdf}) and (\ref{expnint}) in (\ref{exppdf}), it becomes an useful expansion for (\ref{continuous}) when $G(\cdot;\theta)$ has not closed-form given by
\begin{eqnarray}\label{usefulexpansion}
f(x)=\frac{\lambda}{1-e^{-\lambda}}g(x;\theta)\sum_{j,k=0}^\infty\frac{(-\lambda)^j}{j!}c_{j,k}x^{k+jc}.
\end{eqnarray}

Let now $X_1,\ldots,X_n$ be a random sample with pdf in the form (\ref{continuous}) and define $X_{i:n}$ the $i$th order statistic. The pdf of the $X_{i:n}$, say $f_{i:n}$, is given by
\begin{eqnarray}\label{pdforder}
f_{i:n}(x)&=&\frac{1}{B(i,n-i+1)}f(x)F(x)^{i-1}\{1-F(x)\}^{n-i}\nonumber\\
&=&\frac{\lambda g(x;\theta)e^{-\lambda G(x;\theta)}}{B(i,n-i+1)(1-e^{-\lambda})^n}\{1-e^{-\lambda G(x;\theta)}\}^{i-1}\{e^{-\lambda G(x;\theta)}-e^{-\lambda}\}^{n-i}.
\end{eqnarray}   

By using binomial expansion for the terms $\{1-e^{-\lambda G(x;\theta)}\}^{i-1}$ and $\{e^{-\lambda G(x;\theta)}-e^{-\lambda}\}^{n-i}$ in (\ref{pdforder}), it follows 

\begin{eqnarray}\label{exporder}
f_{i:n}(x)&=&\frac{(1-e^{-\lambda})^{-n}}{B(i,n-i+1)}\sum_{j=0}^{i-1}\sum_{k=0}^{n-i}\frac{(-1)^{n+j-k-i}}{j+k+1}\binom{i-1}{j}\binom{n-i}{k}e^{-\lambda(n-k-i)}\nonumber\\
&&\times(1-e^{-\lambda(j+k+1)})f_{j,k}(x),
\end{eqnarray}  
where $f_{j,k}(\cdot)$ denotes the pdf of a random variable with exp-$G$$(\lambda(j+k+1),\theta)$ distribution. Therefore,
the pdf of $X_{i:n}$ can be written as linear combination of pdf's in the form (\ref{continuous}) and, hence, the mathematical properties of the order statistics can be obtained from associated exp-$G$ distribution.\\

We hardly need to emphasize the necessity and importance of moments in any statistical analysis
especially in applied work. Some of the most important features and characteristics of a distribution
can be studied through moments, e.g., tendency, dispersion, skewness and kurtosis. We now give general expressions for the moments of the family exp-$G$ of distributions. \\

Consider $X$ and $Y$ be random variables with exp-$G(\lambda,\theta)$ and $G$ distributions, respectively. When $G(.,\theta)$ has closed-from, an useful expression for the $r$th moment of the exp-$G$ distributions it follows from (\ref{exppdf}) and it is given in function of the probability weighted moments of the $Y$:
\begin{eqnarray}\label{mom1}
E(X^r)=\frac{\lambda}{1-e^{-\lambda}}\sum_{j=0}^\infty \frac{(-\lambda)^j}{j!}E\{Y^rG(Y;\theta)^j\}.
\end{eqnarray}

In particular, formula \eqref{mom1} provides us another proof of condition $v)$ in Proposition \ref{prop1}.

If $G(\cdot,\theta)$ has not closed-from, from (\ref{usefulexpansion}) we obtain the $r$th moment of $X$ in function of the  moments of $Y$:
\begin{eqnarray}\label{mom2}
E(X^r)=\frac{\lambda}{1-e^{-\lambda}}\sum_{j,k=0}^\infty\frac{(-\lambda)^j}{j!}c_{j,k}E(Y^{r+k+jc}).
\end{eqnarray}
In particular, if $c$ is integer non-negative, the moments of $X$ are given in function of the ordinary moments of $Y$. Finally, with the result (\ref{exporder}) the $r$th moment of the $i$th order statistic is given by
\begin{eqnarray}\label{momorder}
E(X_{i:n}^r)&=&\frac{(1-e^{-\lambda})^{-n}}{B(i,n-i+1)}\sum_{j=0}^{i-1}\sum_{k=0}^{n-i}\frac{(-1)^{n+j-k-i}}{j+k+1}\binom{i-1}{j}\binom{n-i}{k}e^{-\lambda(n-k-i)}\nonumber\\
&&\times(1-e^{-\lambda(j+k+1)})E(Z_{j,k}^r),
\end{eqnarray}  
where $Z_{j,k}$ has exp-$G$$(\lambda(j+k+1),\theta)$ distribution. The expansions (\ref{exppdf}), (\ref{usefulexpansion}) and (\ref{exporder}) are main results of this Section and plays an important role in this paper. 

\subsection{Estimation and inference}

Let $X$ be a random variable with exp-$G(\lambda,\theta)$ distribution, with $\lambda\neq0$. The log-density of $X$ with observed value $x$ is given by 
\begin{eqnarray*}
\ell=\ell(\lambda,\theta)=\log\lambda-\log(1-e^{-\lambda})+\log g(x;\theta)-\lambda G(x;\theta)
\end{eqnarray*}
and the associated score function is $U=(\partial\ell/\partial\lambda,\partial\ell/\partial\theta)^\top$, where  
$$\frac{\partial\ell}{\partial\lambda}=\frac{1}{\lambda}-\frac{1}{e^\lambda-1}-G(x;\theta),\quad \frac{\partial\ell}{\partial\theta}=U^{*}(\theta)-\lambda\frac{\partial G(x;\theta)}{\partial\theta},$$
with $U^{*}(\theta)$ being the associated score function of the log-density of a random variable with pdf $g(.,\theta)$.
From regularity conditions, we have $E\{G(X;\theta)\}=\lambda^{-1}-(e^{\lambda}-1)^{-1}$ and $E\{\partial G(X;\theta)/\partial\theta\}=\lambda^{-1}E\{U^{*}(\theta)\}$.\\

The information matrix
$K=K((\lambda,\theta)^\top)$ is
$$K = \left(\begin{array}{ccc}
\kappa_{\lambda,\lambda}&\kappa_{\lambda,\theta}\\
\kappa_{\theta,\lambda}&\kappa_{\theta,\theta}\\
\end{array}\right),$$
where 
$$\kappa_{\lambda,\lambda}=\frac{1}{\lambda^2}-\frac{e^\lambda}{(e^\lambda-1)^2},\,\, \kappa_{\theta,\theta}=\lambda E\left\{\frac{\partial^2G(X;\theta)}{\partial\theta\partial\theta^\top}\right\}-E\left\{\frac{\partial U^{*}(\theta)}{\partial\theta} \right\},\,\, \kappa_{\theta,\lambda}=\lambda^{-1}E\{U^{*}(\theta)\}.$$

For a random sample $x=(x_1,\ldots,x_n)$ of size $n$ from $X$ and $\Theta = (\lambda,\theta)^T$, the total log-likelihood is
$$\ell_n=\ell_n(\Theta)=\sum_{i=1}^n \ell^{(i)},$$ where $\ell^{(i)}$ is
the log-likelihood for the $i$th observation ($i=1,\ldots,n$) as given before.
The total score function is $U_n=U_n(\Theta)=\sum_{i=1}^n U^{(i)}$, where $U^{(i)}$
for $i=1,\ldots,n$ has the form given earlier and the total information matrix is
$K_n(\theta)=n K(\Theta)$.

The maximum likelihood estimator (MLE) $\hat\Theta$ of $\Theta$ is obtained numerically from the solution of
the non-linear system of equations $U_n=0$. Under conditions that are fulfilled for
the parameter $\Theta$ in the interior of the parameter space but not on the boundary,
the asymptotic distribution of
\begin{eqnarray*}
\sqrt n (\hat\Theta-\Theta)\,\,\,\,\stackrel{A}{\sim}\,\,\,\,N_{k+1}(0,K(\Theta)^{-1}),
\end{eqnarray*}
where `$\stackrel{A}{\sim}$' stands for the asymptotic distribution.
The asymptotic multivariate normal $N_{k+1}(0,K_n(\hat\Theta)^{-1})$ distribution of
$\hat\Theta$ can be used to cons\-truct approximate confidence regions for some
parameters and for the hazard and survival functions. In fact, an $100(1-\gamma)\%$
asymptotic confidence interval for each parameter $\Theta_i$ is given by
$$ACI_i=(\hat\Theta_i-z_{\gamma/2}\sqrt{\hat\kappa^{\Theta_i,\Theta_i}},\hat{\Theta_i}
+z_{\gamma/2}\sqrt{\hat\kappa^{\Theta_i,\Theta_i}}),$$
where $\hat\kappa^{\Theta_i,\Theta_i}$ denotes the $i$th diagonal element of
$K_n(\hat\Theta)^{-1}$ for $i=1,\ldots,k+1$ and $z_{\gamma/2}$ is the quantile $1-\gamma/2$
of the standard normal distribution. The asymptotic normality is also useful
for testing goodness of fit of the exp-$G$ distribution and for comparing 
this distribution with some of its special submodels using one of the three well-known asymptotically
equivalent test statistics - namely, the likelihood ratio (LR)
statistic, Rao ($S_R$) and Wald ($W$) statistics. Consider the partition
$\Theta=(\Theta_1^T,\Theta_2^T)^T$ of the vector of parameters for the
exp-Weibull distribution. The total score function $U_n=(U_1^{T},U_2^{T})^{T}$
and the total Fisher information matrix and its inverse
$$K_n = \left(\begin{array}{cc}
K_{11}&K_{12}\\
K_{21}&K_{22}
\end{array}\right)
, \,\,
K_n^{-1} = \left(\begin{array}{cc}
K^{11}&K^{12}\\
K^{21}&K^{22}
\end{array}\right),$$
are assumed partitioned in the same way as $\Theta$.
The LR statistic for testing the null hypothesis $H_0:\Theta_1 =\Theta_1^{(0)}$
versus the alternative hypothesis $H_A:\Theta_1 \neq \Theta_1^{(0)}$ is given
by $w= 2\{\ell(\hat{\Theta})-\ell(\tilde{\Theta})\}$, where
$\hat\Theta$ and $\tilde\Theta$ denote the MLEs under the null and the
alternative hypotheses, respectively. The statistic $w$ is asymptotically (as $n\to\infty$)
distributed as $\chi_q^2$, where $q$ is the dimension of the vector $\theta_1$
of interest. The score statistic for testing $H_0$ is $S_R = \widetilde{U}_1^T \widetilde{K}^{11}\widetilde{U}_1$,
where $\widetilde{U}_1$ and $\widetilde{K}^{11}$ are the components of $U_n$ and $K_n^{-1}$
corresponding to $\Theta_1$ evaluated at $\tilde\Theta$. The score statistic $S_R$
has asymptotically the $\chi_q^2$ distribution and has an advantage over the LR
since it only needs the estimation under the null hypothesis but requires the
inverse Fisher information matrix. The Wald statistic for testing the null hypothesis $H_0:\Theta_1 =\Theta_1^{(0)}$
is given by $W=(\hat\Theta_1-\Theta_1^{(0)})^T \widehat{K}^{11(-1)}(\hat{\Theta}_1-\Theta_1^{(0)})$,
where $\widehat{K}^{11}$ is the component of the inverse information matrix $K_n^{-1}$
corresponding to $\Theta_1$ evaluated at $\hat\Theta$. The Wald statistic
$W$ has also under $H_0$ an asymptotic $\chi_q^2$ distribution. The Wald and
score statistics are very used in practice and our derivation of the
information matrix will be very convenient in modelling the exp-$G$
distributions.
\subsection{Modified profile likelihood estimator}

Since $\lambda$ is a parameter added to some distribution, it can be seen as a nuisance parameter. With this in mind, we will advance a modified profile estimator for $\theta$. From the last subsection, we have that
$$\frac{\partial\ell}{\partial\theta}=U^{*}(\theta)-\lambda\frac{\partial G(x;\theta)}{\partial\theta},$$
with 
$$E\left(\frac{\partial\ell}{\partial\theta}\right) = 0.$$
Therefore, if $\partial G(x;\theta)/\partial\theta^\top \partial G(x;\theta)/\partial\theta$, that belongs to $\mathbb{R}$, does not vanish for all values in some open neighbourhood of the true value of $\theta$, let
$$\breve{\lambda} = \left\{\frac{\partial G(x;\theta)}{\partial\theta^\top}\frac{\partial G(x;\theta)}{\partial\theta}\right\}^{-1}{\frac{\partial G(x;\theta)}{\partial\theta^\top} U^*(\theta)},$$
with
\begin{eqnarray*}
\frac{\partial\breve{\lambda}}{\partial\theta} &=& -2\left\{\frac{\partial G(x;\theta)}{\partial\theta^\top}\frac{\partial G(x;\theta)}{\partial\theta}\right\}^{-2}\left\{\frac{\partial^2G(x;\theta)}{\partial\theta^\top\partial\theta^\top} \frac{\partial G(x;\theta)}{\partial\theta}\right\}{\frac{\partial G(x;\theta)}{\partial\theta^\top} U^*(\theta)}\\
&+&\left\{\frac{\partial G(x;\theta)}{\partial\theta^\top}\frac{\partial G(x;\theta)}{\partial\theta}\right\}^{-1}\left\{\frac{\partial^2 G(x;\theta)}{\partial\theta^\top\partial\theta^\top} U^*(\theta) + \frac{\partial G(x;\theta)}{\partial\theta^\top}\frac{\partial U^*(\theta)}{\partial\theta} \right\},
\end{eqnarray*}
where $\partial^2 G(x;\theta)/\partial\theta^\top\partial\theta^\top$ stands for the row vector containing the diagonal elements of the Hessian matrix of $G$, $\partial^2 G(x;\theta)/\partial\theta\partial\theta^\top$, and $\partial U^*(\theta)/\partial\theta$ stands for the column vector $(\partial U_1^*(\theta)/\partial\theta_1,\ldots,\partial U_k^*(\theta)/\partial\theta_k)^\top$.

We, therefore, obtain the modified profile likelihood function:
$$\breve{\ell}=\breve{\ell}(\theta)=\log\breve{\lambda}-\log(1-e^{-\breve{\lambda}})+\log g(x;\theta)-\breve{\lambda} G(x;\theta).$$

The modified profile estimator for $\theta$ can be obtained by maximizing $\breve{\ell}$. Let $V$ be the estimating equation given by
$$V(\theta) = \frac{\partial\breve{\ell}}{\partial{\theta}} = \frac{1}{\breve{\lambda}}\frac{\partial\breve{\lambda}}{\partial\theta} - \frac{1}{e^{\breve{\lambda}}-1}\frac{\partial\breve{\lambda}}{\partial\theta} + U^*(\theta) - \frac{\partial\breve{\lambda}}{\partial\theta} G(x;\theta) - \breve{\lambda}\frac{\partial G(x;\theta)}{\partial\theta},$$
one may also obtain the profile likelihood estimator by solving the equation $V_n(\theta) = \sum_{i=1}^n V^{(i)}(\theta) = 0$.

\subsection{Interest case $\lambda=0$}

We now discuss estimation and inference when $\lambda=0$. It is very important to discuss this case because we are interested in testing the hypotheses $H_0: \lambda=0$ versus $H_1: \lambda\neq0$, i.e., to test if the exp-$G$ fit is significantly better than $G$ fit. The next result plays a important role in this paper.\\

\begin{theorem}\label{teo1}
Let $F_\lambda(\cdot)$ and $f_\lambda(\cdot)$ be the cdf and pdf defined by (\ref{expg}) and (\ref{continuous}), respectively. The following conditions are true:\\
\begin{itemize}
\item[i)] If $G$ is continuous then $F_\lambda\rightarrow G$ uniformly when $\lambda\rightarrow0$;
\item[ii)] $f_\lambda\rightarrow g$ uniformly when $\lambda\rightarrow0$, consequently $\ell_n(\lambda,\theta)\rightarrow\ell^*_n(\theta)$, where $\ell^*_n(\theta)$ is the log-likelihood associated to $G$;
\item[iii)] $\partial\ell/\partial\lambda\rightarrow1/2-G(x;\theta)$ and $\partial\ell/\partial\theta\rightarrow U^*(\theta)$, when $\lambda\rightarrow0$;
\item[iv)] $\kappa_{\lambda,\lambda}\rightarrow1/12$, $\kappa_{\theta,\theta}\rightarrow\int \partial U^*(\theta)/\partial\theta\partial\theta^\top dG$ and $\kappa_{\lambda,\theta}\rightarrow-(1/2,1/2,1/2)^\top$, when $\lambda\rightarrow0$, with $\int \partial U^*(\theta)/\partial\theta\partial\theta^\top dG$ being the information matrix with respect to $G$;
\item[v)] If $G$ is regular and $(\lambda_0,\theta_0)\in\Theta$, then
$$\sqrt{n}\{(\hat{\lambda},\hat\theta^\top)^\top-(\lambda_0,\theta_0^\top)^\top\}\stackrel{d}{\rightarrow}N_{p+1}(0_{q+1},K(\lambda_0,\theta_0)^{-1});$$
\item[vi)] If $G$ is regular and $(\lambda_0,\theta_0)\in\Theta$, then the likelihood ratio, Wald and Score statistics has null asymptotic distribution $\chi^2_q$, where $q$ is the number of parameters estimated in alternative hypothesis minus the number of parameters estimated in null hypothesis.
\end{itemize}
\end{theorem}

\section{The exp-Weibull distribution}\label{sec3}

We now move to the class of distributions exp-$G$, when $G$ is the cdf of the Weibull distribution, we will call this class of distributions by
exp-Weibull. More precisely, to obtain the exp-Weibull distribution we put in (\ref{expg}) the cdf of the Weibull distribution $G(x)=1-\exp\{-(x/\beta)^\alpha\}$, where $\beta>0$, $\alpha>0$ and $x>0$. Therefore, the cdf of the exp-Weibull distribution given by
\begin{eqnarray*}
F(x)=\frac{1-\exp\{-\lambda(1-e^{-(x/\beta)^\alpha})\}}{1-e^{-\lambda}}, \quad x>0.
\end{eqnarray*}

From the general expressions (\ref{continuous}) and (\ref{haz}) we obtain that the pdf and hazard functions are given by
\begin{eqnarray}\label{pdfexpw}
f(x)=\frac{\lambda\alpha\beta^{-\alpha}}{1-e^{-\lambda}}x^{\alpha-1}\exp\{-\lambda(1-e^{-(x/\beta)^\alpha})-(x/\beta)^\alpha\}, \quad x>0
\end{eqnarray}
and 
\begin{eqnarray*}
h(x)=\frac{\lambda\alpha\beta^{-\alpha} x^{\alpha-1}e^{-(x/\beta)^\alpha}}{1-\exp\{-\lambda e^{-(x/\beta)^\alpha}\}}, \quad x>0,
\end{eqnarray*}
respectively. \\

\begin{figure}[h!]
	\centering\includegraphics[width=0.45\textwidth]{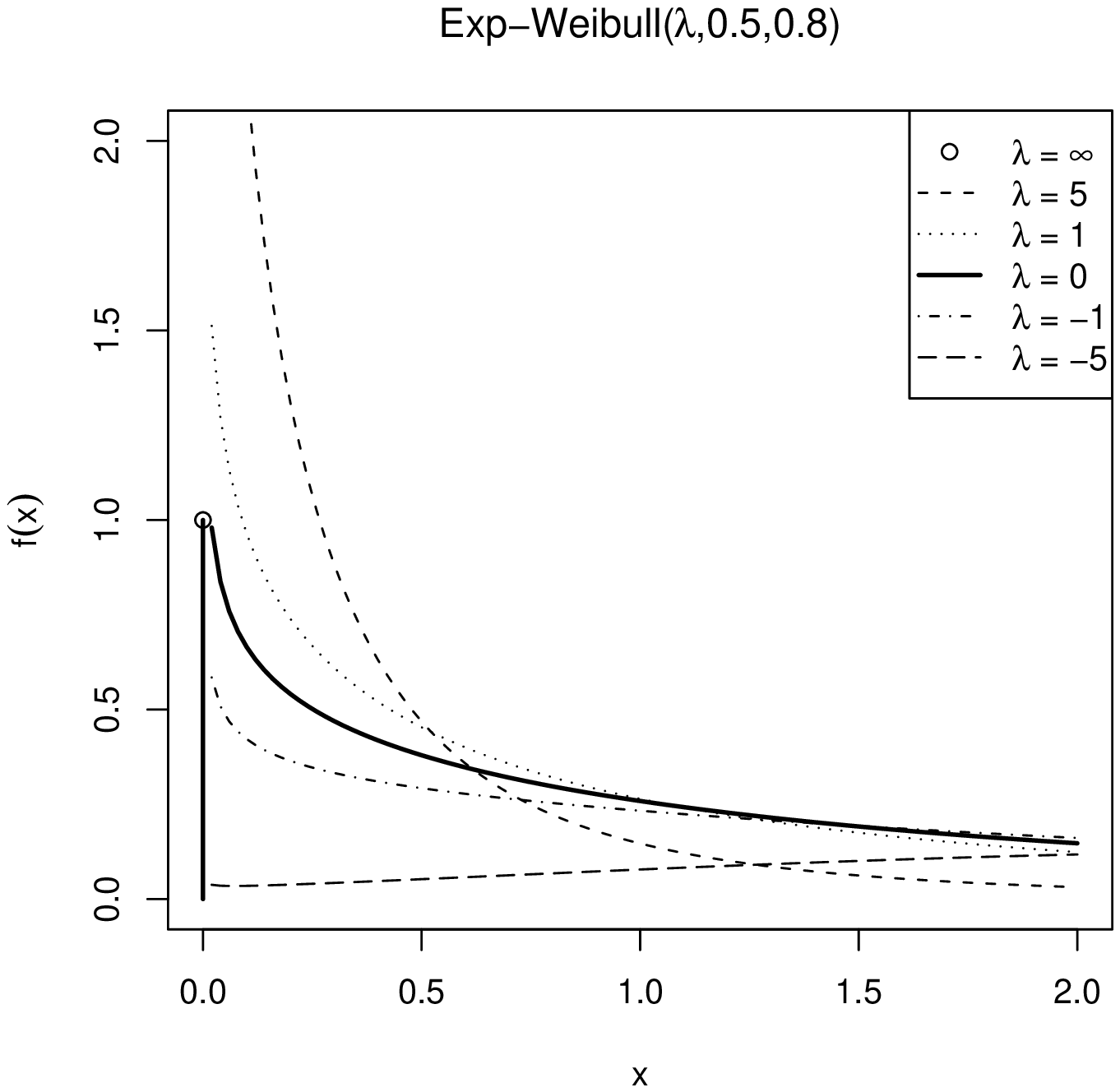}\includegraphics[width=0.45\textwidth]{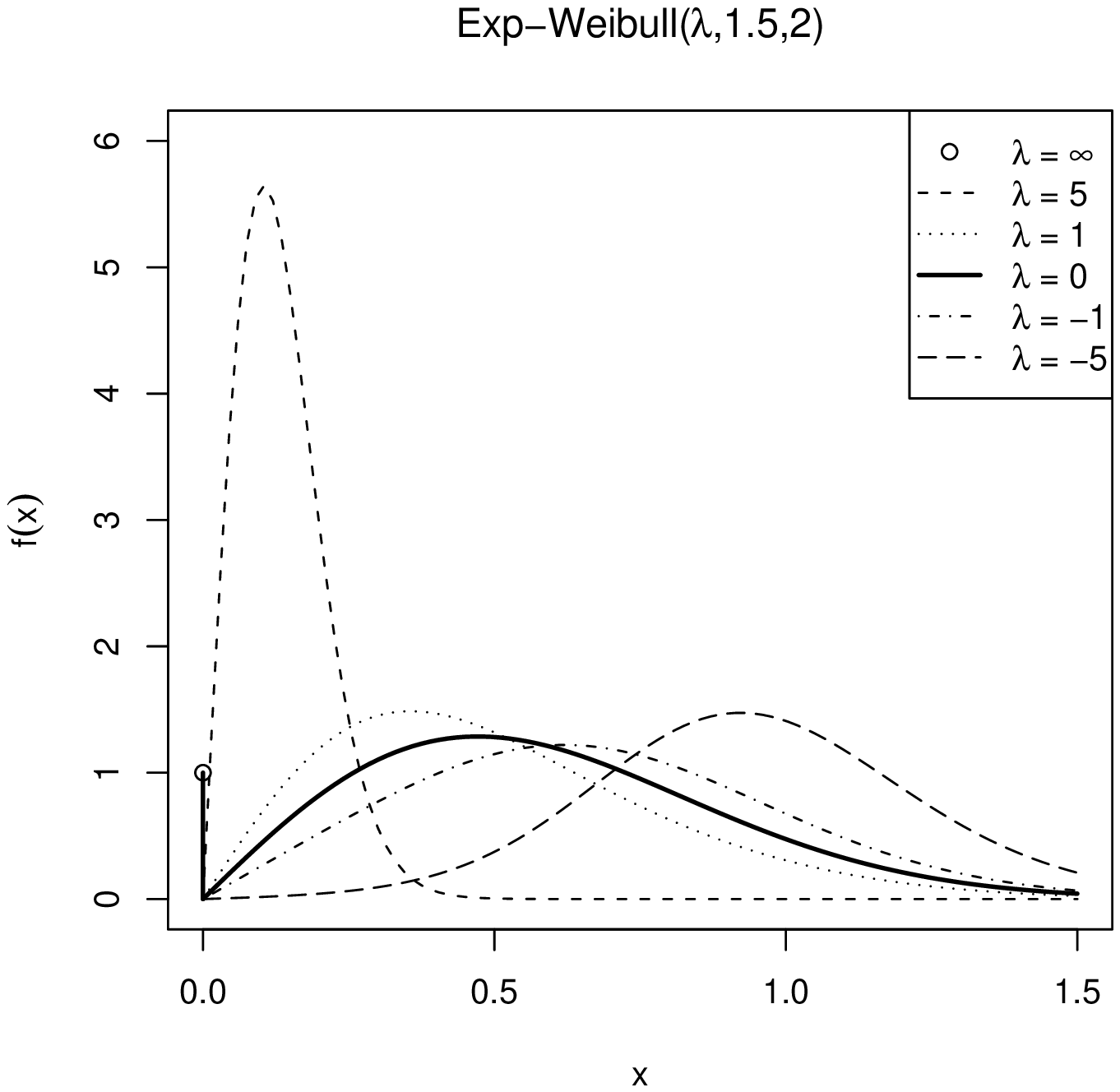}
	\includegraphics[width=0.45\textwidth]{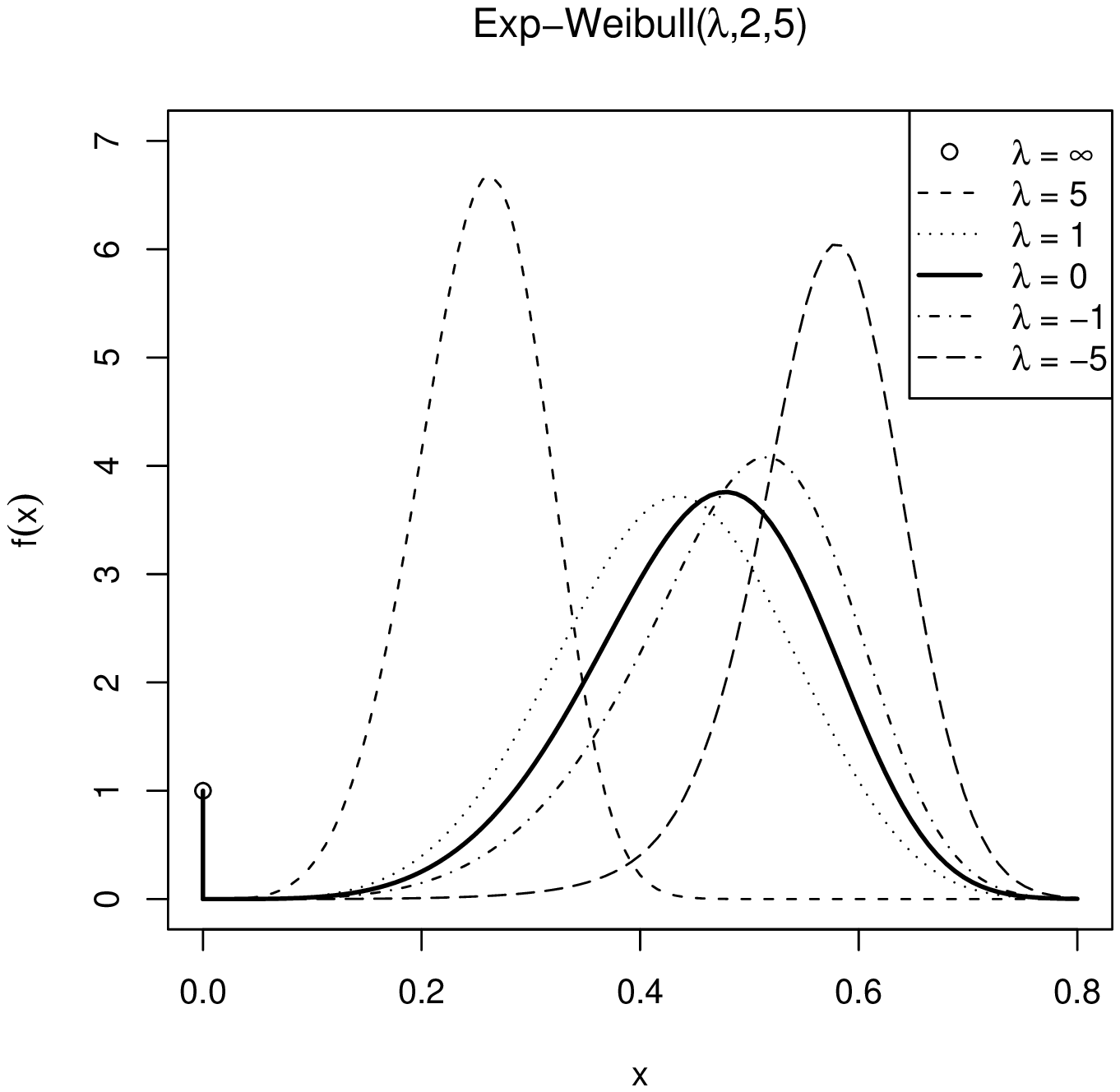}
	\caption{Graphics of the pdf of the exp-Weibull distribution for some values of the parameters.}
	\label{pdf}
\end{figure}

\begin{figure}[h!]
	\centering\includegraphics[width=0.45\textwidth]{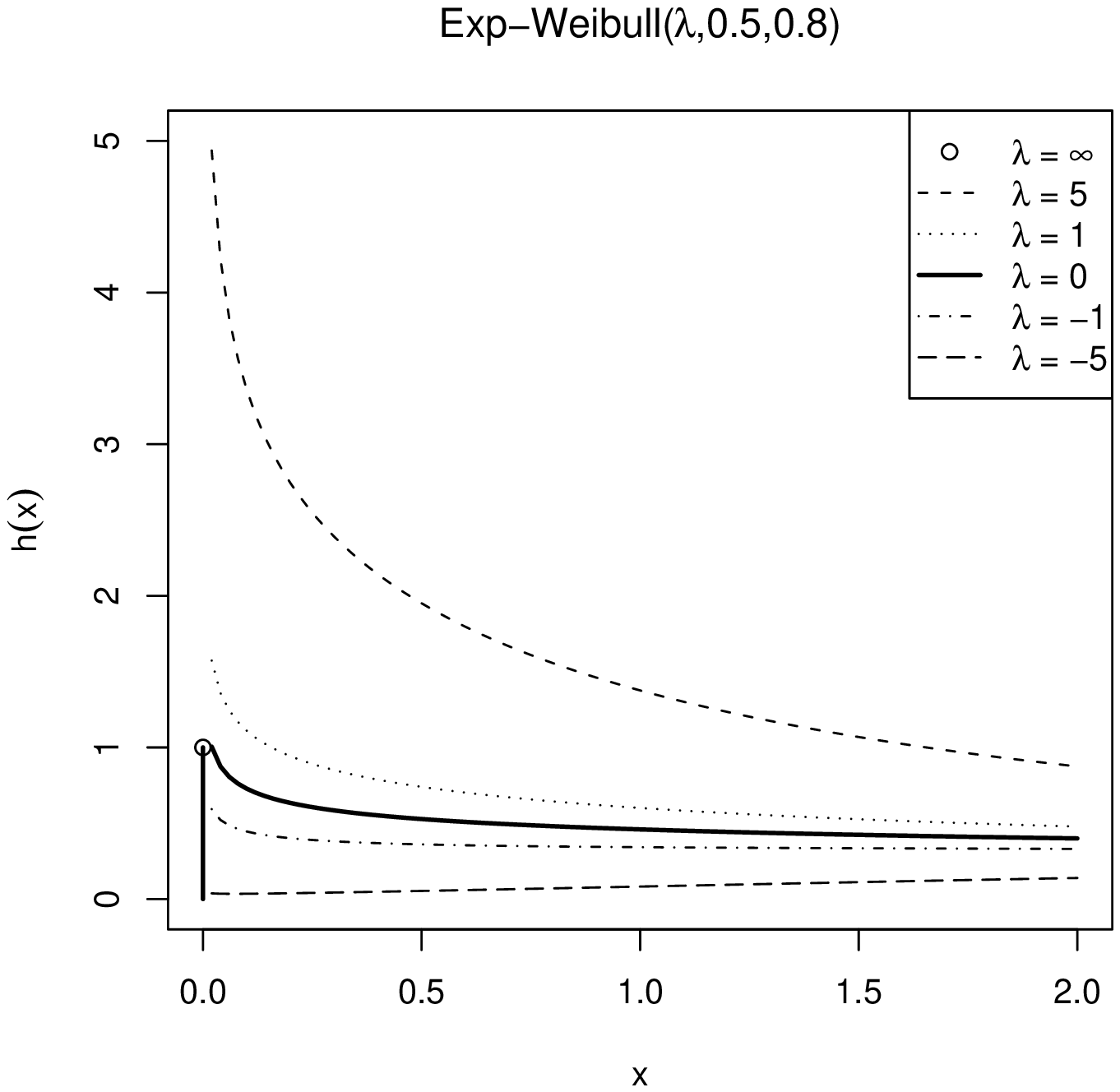}\includegraphics[width=0.45\textwidth]{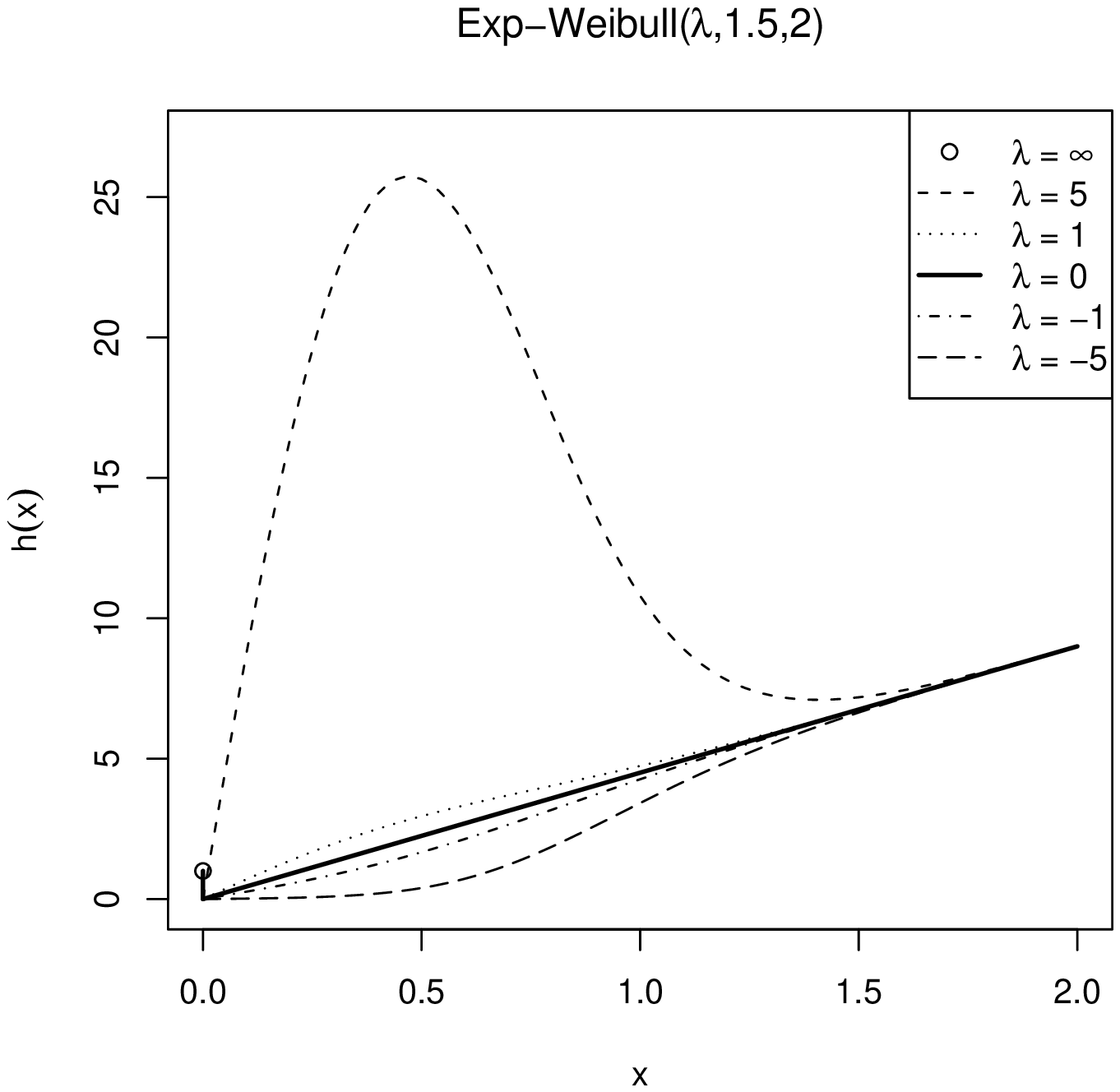}
	\includegraphics[width=0.45\textwidth]{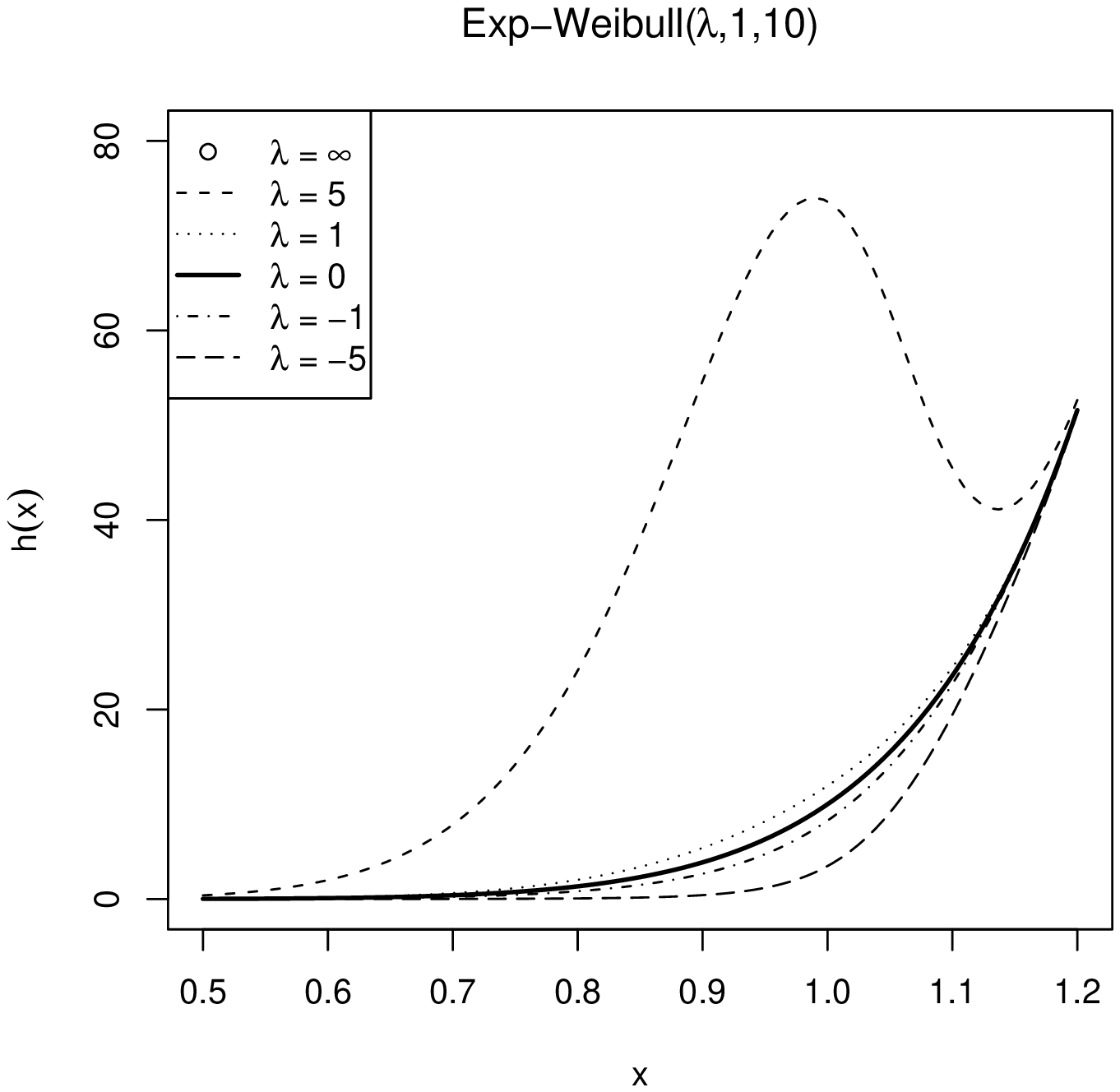}
				\caption{Graphics of the hazard function of the exp-Weibull distribution for some values of the parameters.}
\label{hazard}
\end{figure}

We now illustrate the flexibility of this class of distributions by presenting some graphics of both the pdf and hazard functions. Figure \ref{pdf} shows the plots of the pdf of the exp-Weibull distribution for some values of $\alpha$ and $\beta$, and for $\lambda=-5,-1,0, 1, 5, \infty$. We note that when the the value of $\lambda$ increases the pdf becomes more `peaked'. Figure \ref{hazard} contains the plots of hazard function of the exp-Weibull distribution for different values of $\alpha$ and $\beta$ and $\lambda=-5,-1,0, 1, 5, \infty$. We note that the behaviour of the hazard function of the Weibull distribution is close to the behaviour of the graphics with $\lambda=1.0$, and as the value of $\lambda$ increases, the behaviour of the hazard function of the exp-Weibull becomes very different from the behaviour of the hazard function of the Weibull distribution, showing that as the value of $\lambda$ gets larger the exp-Weibull ``moves away'' from the Weibull distribution, and gets closer to the Dirac mass at zero, as remarked on the end of the last Section.\\

\subsection{Order statistics and moments}
The pdf of the $i$th order statistic of a random sample from exp-$W(\lambda,\beta,\alpha)$ distribution is given by
\begin{eqnarray*}
f_{i:n}(x)&=&\frac{\lambda\alpha\beta^{-\alpha} x^{\alpha-1}\exp\left\{-\lambda(1-e^{-(x/\beta)^\alpha})-( x/\beta)^\alpha\right\}}{B(i,n-i+1)(1-e^{-\lambda})^n}\left\{1-e^{-\lambda(1-e^{-(x/\beta)^\alpha})}\right\}^{i-1}\times\\
&&\left\{e^{-\lambda(1-e^{-(x/\beta)^\alpha})}-e^{-\lambda}\right\}^{n-i}, \quad x>0.
\end{eqnarray*}

We will now obtain series representation for the moments of the exp-Weibull distribution and of the order statistics. To this end, let $X$ be a random variable following a exp-Weibull distribution with parameters $\beta>0$, $\alpha>0$ and $\lambda>0$. From now on we will use the notation $X\sim \mbox{exp-Weibull}(\lambda,\beta,\alpha)$ to indicate this fact.\\

We have the probability weighted moment of a random variable $Y$ following Weibull distribution with parameter vector $\theta=(\beta,\alpha)^\top$ can be written as 
$E\{Y^r G(Y;\theta)^j\}=\beta^{r}\int_0^1(-\log u)^{r/\alpha}(1-u)^jdu$.
 Therefore, from (\ref{mom1}) it follows that the $r$th moment of $X$ is
\begin{eqnarray}\label{momentsew1}
E(X^r)=\frac{\lambda\beta^r}{1-e^{-\lambda}}\sum_{k=0}^\infty \frac{(-\lambda)^k}{k!}\int_0^1(1-u)^k(-\log u)^{r/\alpha}du.
\end{eqnarray}
We now give an alternative expression to (\ref{momentsew1}) more simple. The $r$th moment of $X$ is
\begin{eqnarray*}
E(X^r)=\int_0^\infty x^rf(x)dx=\int_0^\infty  \frac{\lambda\alpha\beta^{-\alpha}}{1-e^{-\lambda}}x^{r+\alpha-1}\exp\{-\lambda(1-e^{-(x/\beta)^\alpha})-(x/\beta)^\alpha\}dx.
\end{eqnarray*}
Now, expading $\exp\{\lambda e^{-(x/\beta)^\alpha}\}$ in Taylor's series we get
\begin{eqnarray*}
E(X^r)&=&\frac{\lambda e^{-\lambda}}{1-e^{-\lambda}}\sum_{k=0}^\infty \frac{\lambda^k}{k!}\int_0^\infty x^r \alpha\beta^{-\alpha} x^{\alpha-1}e^{-\{(k+1)^{1/\alpha}x/\beta\}^\alpha} dx=\frac{\lambda e^{-\lambda}}{1-e^{-\lambda}}\sum_{k=0}^\infty \frac{\lambda^kE(Y_k^r)}{(k+1)!},
\end{eqnarray*}
where $Y_k$ follows the Weibull distribution with parameters $(k+1)^{1/\alpha}/\beta$ and $\alpha$, and the interchange between the series and integral being possible due to Fubini's theorem together with the fact that we are dealing with positive integrand. Hence, we have that the $r$th moment of a exp-Weibull distribution can be written as
\begin{eqnarray}\label{momentsew}
E(X^r)=\lambda\beta^r\frac{\Gamma(r/\alpha+1)}{e^\lambda-1}\sum_{k=0}^\infty \frac{\lambda^k}{k!(k+1)^{r/\alpha+1}}.
\end{eqnarray}

\begin{figure}[h!]
	\centering
	\includegraphics[width=0.45\textwidth]{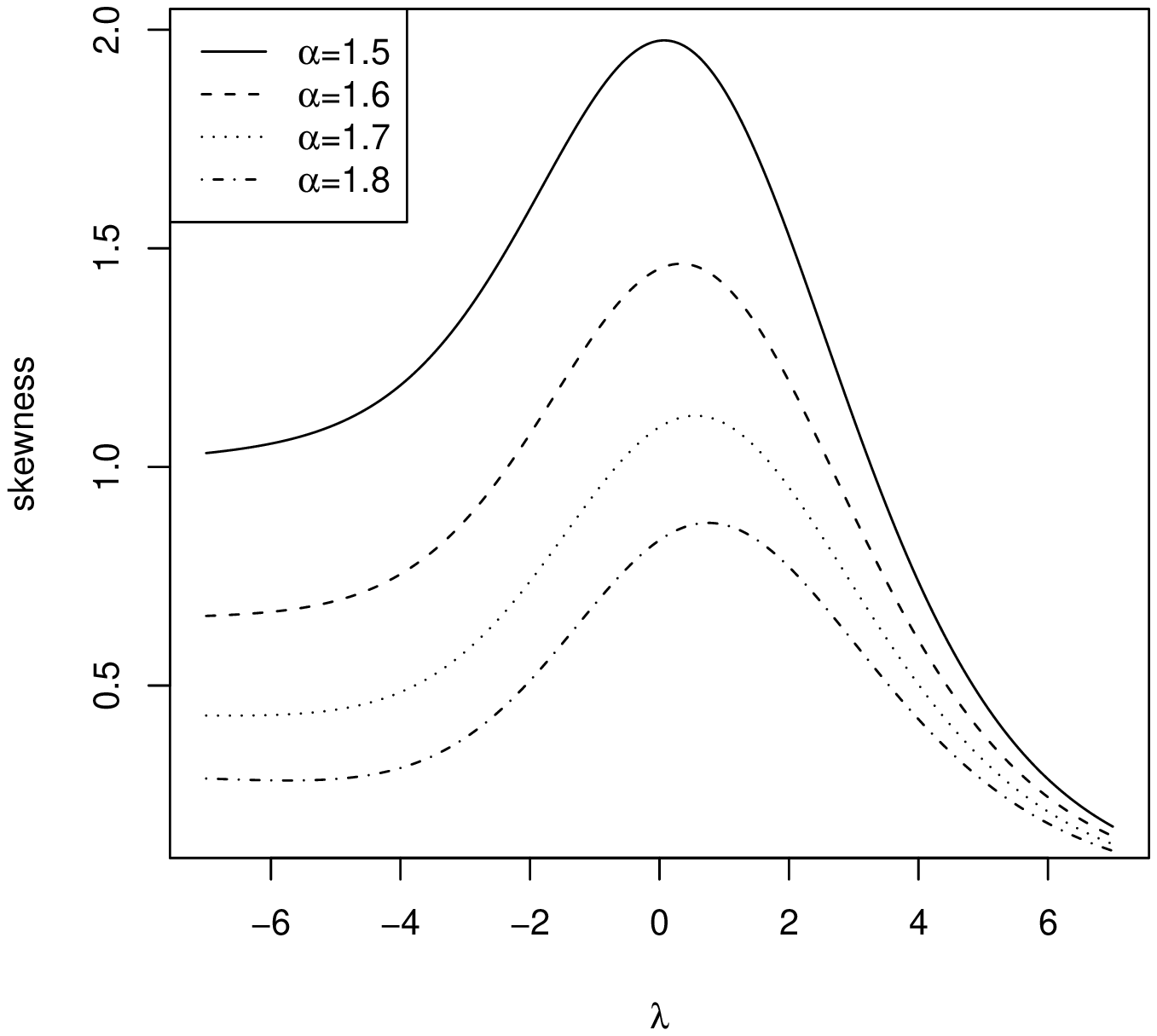}\includegraphics[width=0.45\textwidth]{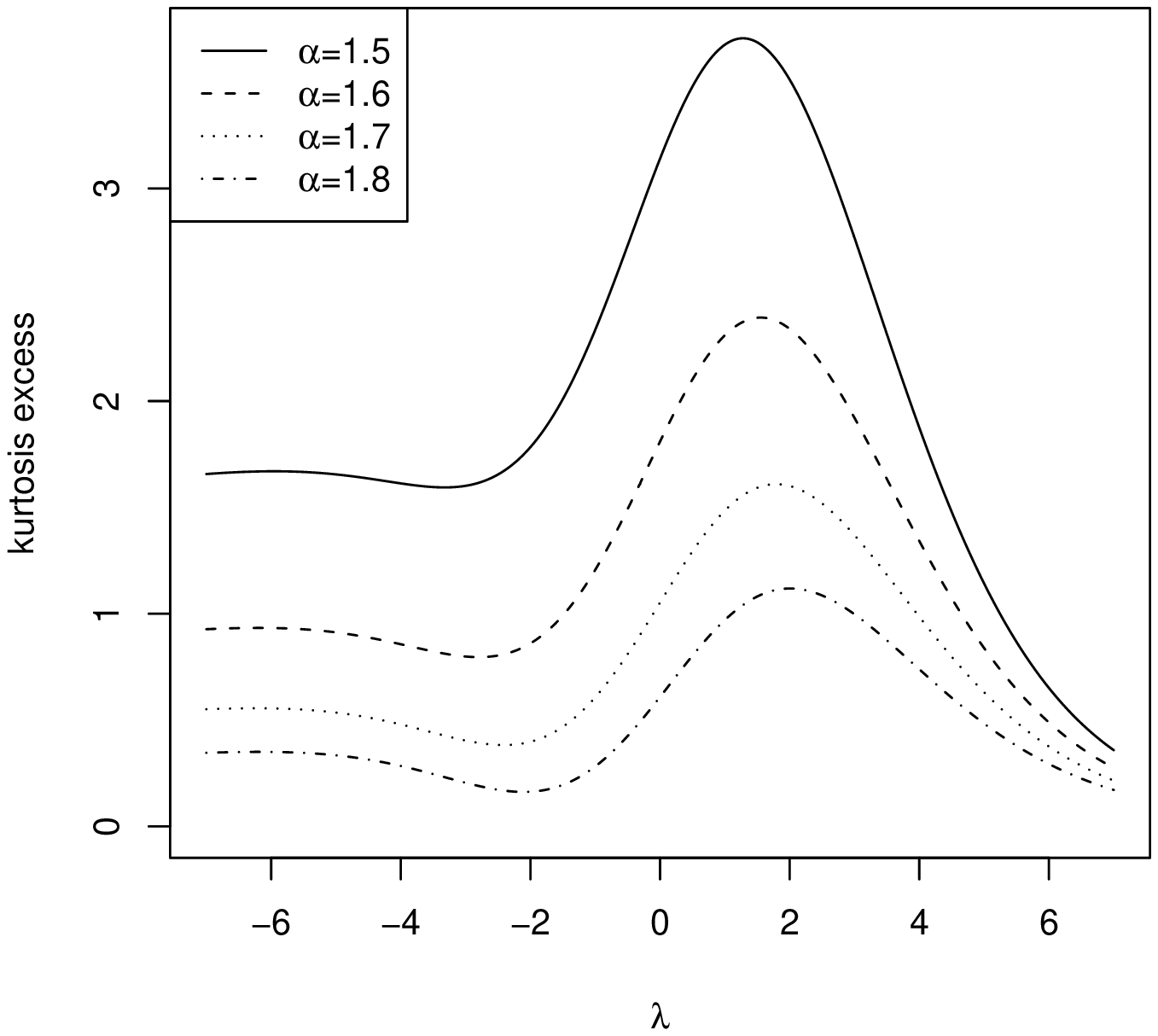}
			\caption{Skewness and kurtosis of the exp-Weibull distribution for some values of the parameters.}
			\label{skewkur}
\end{figure}

Figure \ref{skewkur} shows skewness and kurtosis of the exp-Weibull distribution, obtained from application of the formula of the moments above, for $\beta=0.5$ and some values of $\alpha$ as function of $\lambda$. We now note from (\ref{momentsew}) that all moments of the exp-Weibull distribution tends to zero as $\lambda$ increases to infinity, which is a very remarkable fact. So, as we can note from Figure \ref{skewkur}, as $\lambda$ increases, the skewness tends to zero, as well as the kurtosis, one more time reflecting the expected behaviour of the limiting distribution as $\lambda\to\infty$.  

An expression for the $r$th moment of the $i$th order statistic of the exp-Weibull distribution, say $X_{i:n}$, follows from (\ref{momorder}) and (\ref{momentsew}):
\begin{eqnarray}\label{momorderew}
E(X_{i:n}^r)&=&\frac{\lambda\beta^{r}\Gamma(r/\alpha+1)}{B(i,n-i+1)(1-e^{-\lambda})^{n}}\sum_{l=0}^\infty\sum_{j=0}^{i-1}\sum_{k=0}^{n-i}(-1)^{n+j-k-i}\binom{i-1}{j}\binom{n-i}{k}\times\nonumber\\
&&e^{-\lambda(n+j-i+1)}\frac{\{\lambda(j+k+1)\}^l}{l!(l+1)^{r/\alpha+1}}.
\end{eqnarray}
Expressions (\ref{momentsew1}) and (\ref{momorderew}) show the importance of the expansions given in Subsection (\ref{expansions}). Furthermore, result (\ref{momentsew}) shows that alternative expressions to (\ref{mom1}) and (\ref{mom2}) can be obtained depending of the $G$ distribution. 

\subsection{Order statistics and moments of the exp-Fréchet distribution}

In this brief subsection we use the reciprocal property of the exp-$G$ distributions to obtain expressions for the moments and order statistics of the exp-Fréchet distribution.

Let $Y\sim$ exp-Fr($\lambda,\beta,\alpha$) and $Y_{i:n}$ be the $i$th order statistics from a random sample, of size $n$, of the exp-Fréchet distribution. From formulae \eqref{momentsew} and \eqref{momorderew}, we have that the moments of $Y$ and $Y_{i:n}$ are
$$E(Y^r)=-\lambda\frac{\Gamma(1-r/\alpha)}{\beta^{r}(e^{-\lambda}-1)}\sum_{k=0}^\infty \frac{(-\lambda)^k}{k!(k+1)^{1-r/\alpha}}$$
and
\begin{eqnarray*}
E(X_{i:n}^r)&=&\frac{-\lambda\Gamma(1-r/\alpha)}{\beta^{r}B(i,n-i+1)(1-e^{\lambda})^{n}}\sum_{l=0}^\infty\sum_{j=0}^{i-1}\sum_{k=0}^{n-i}(-1)^{n+j-k-i}\binom{i-1}{j}\binom{n-i}{k}\times\nonumber\\
&&e^{\lambda(n+j-i+1)}\frac{\{-\lambda(j+k+1)\}^l}{l!(l+1)^{1-r/\alpha}},
\end{eqnarray*}
respectively, for $r<\alpha$.
\subsection{Score function and information matrix}

Let $\theta=(\lambda,\beta,\alpha)^T$ be the parameter vector and $X$ random variable with $\mbox{exp-Weibull}(\lambda,\beta,\alpha)$ distribution. The log-density $\ell=\ell(\theta)$ for the random variable $X$ with observed value $x$ is given by
\begin{eqnarray*}
\ell= - \alpha \log \beta + \log (\alpha \lambda )-{\left( \frac{x}{\beta } \right) }^{\alpha } - \lambda\{1-e^{-{\left( \frac{x}{\beta } \right) }^{\alpha }}\} - \log (1 - e^{-\lambda }) + 
  \left( \alpha-1  \right) \log x, \quad x>0.
\end{eqnarray*}

The score function is given by
\begin{eqnarray*}
\frac{\partial l}{\partial \lambda} &=&-1 + e^{-{\left( \frac{x}{\beta } \right) }^{\alpha }} + \frac{1}{1 - e^{\lambda }} + \frac{1}{\lambda } \\
\frac{\partial l}{\partial \beta} &=& \alpha\,\beta^{-1} \,\left[ -1 + {\left( \frac{x}{\beta } \right) }^{\alpha }\left\{1 + \lambda \,\ e^{-{\left( \frac{x}{\beta } \right) }^{\alpha }} \right\}\right] \\
\frac{\partial l}{\partial \alpha} &=& \frac{1}{\alpha } + \log (x)- \log (\beta ) - {\left( \frac{x}{\beta } \right) }^{\alpha }\,\log \left(\frac{x}{\beta }\right)  \, \left\{1 + \lambda e^{-{\left( \frac{x}{\beta } \right) }^{\alpha }}  \right\} 
\end{eqnarray*}

From the regularity conditions one obtains the following closed-form expressions
$$E\left[e^{-{\left( \frac{X}{\beta } \right) }^{\alpha }}\right] = 1 - \frac{1}{1 - e^{\lambda }} - \frac{1}{\lambda },$$
$$E\left[\left( \frac{X}{\beta } \right)^{\alpha }\left\{1 + \lambda \,\ e^{-{\left( \frac{X}{\beta } \right) }^{\alpha }} \right\}\right] = 1$$
and
$$E\left[{\left( \frac{X}{\beta } \right) }^{\alpha }\,\log \left(\frac{X}{\beta }\right)  \, \left\{1 + \lambda e^{-{\left( \frac{X}{\beta } \right) }^{\alpha }}  \right\}  \right] = \frac{1}{\alpha }- \log (\beta ) + E\{\log (X)\}. $$
For interval estimation and hypothesis tests on the model parameters, we require the
information matrix. We will, therefore, use some of the expressions above to obtain the Fisher's information matrix.  The $3\times 3$ unit information matrix
$K=K((\lambda,\beta,\alpha)^T)$ is
$$K = \left(\begin{array}{ccc}
\kappa_{\lambda,\lambda}&\kappa_{\lambda,\beta}&\kappa_{\lambda,\alpha}\\
\kappa_{\lambda,\beta}&\kappa_{\beta,\beta}&\kappa_{\beta,\alpha}\\
\kappa_{\lambda,\alpha}&\kappa_{\beta,\alpha}&\kappa_{\alpha,\alpha}\\
\end{array}\right),$$
whose elements are
$$\kappa_{\lambda,\lambda} = {\lambda }^{-2}-\frac{e^{\lambda }}{{\left( -1 + e^{\lambda } \right) }^2},\quad\kappa_{\lambda,\beta} = \frac{\alpha}{\beta\lambda} \left[E\left\{\left(\frac{X}{\beta}\right)^\alpha\right\}-1\right],$$
$$\kappa_{\lambda,\alpha} = \frac{1}{\lambda}\left[E\left\{{\left( \frac{X}{\beta } \right) }^{\alpha }\,\log \left(\frac{X}{\beta }\right)\right\} -\frac{1}{\alpha }+ \log (\beta ) - E\{\log (X)\}\right],$$
$$\kappa_{\beta,\beta} = \frac{\alpha}{\beta^3}\left[1 -\alpha\lambda E\left\{e^{-{\left( \frac{X}{\beta } \right) }^{\alpha }}\left( \frac{X}{\beta } \right)^{2\alpha }\right\}\right],$$
$$\kappa_{\beta,\alpha} = \alpha\left[ \lambda E\left\{e^{-{\left( \frac{X}{\beta } \right) }^{\alpha }}\,{\left( \frac{X}{\beta } \right) }^{2\,\alpha }\, \,\log \left(\frac{X}{\beta }\right) \right\} - \frac{1}{\alpha} + \log(\beta) - E\{\log(X)\}\right]$$
and
$$\kappa_{\alpha,\alpha} = \frac{1}{\alpha^2} + E\left[\left(\frac{X}{\beta}\right)^{\alpha} \log\left(\frac{X}{\beta}\right)^{2} 
\left\{1 - \lambda e^{-\left(\frac{X}{\beta}\right)^{\alpha}}  \left(\frac{X}{\beta}\right)^{\alpha} +\lambda e^{-\left(\frac{X}{\beta}\right)^{\alpha}}\right\} \right].$$
These elements of the information matrix depend on some expectations that can be easily obtained through numerical integration. 

\section{The exp-beta distribution}\label{sec4}

Let $Y$ be a random variable following standard beta distribution with parameters $a>0$ and $b>0$. The cdf of $Y$ is given by $G(x;(a,b)^\top)=I_x(a,b)$, where $I_x(a,b)=B(a,b)^{-1}\int_0^x t^{a-1}(1-t)^{b-1}dt$ denotes the incomplete beta function and $B(a,b)=\int_0^1 t^{a-1}(1-t)^{b-1}dt$ is the beta function.  The exp-beta distribution is introduced by taking $G$ as being the cdf of $Y$ in (\ref{expg}). We will denote a random variable $X$ with exp-beta distribution by $X\sim \mbox{exp-beta}(\lambda,a,b)$. \\

\begin{figure}[h!]
	\centering\includegraphics[width=0.45\textwidth]{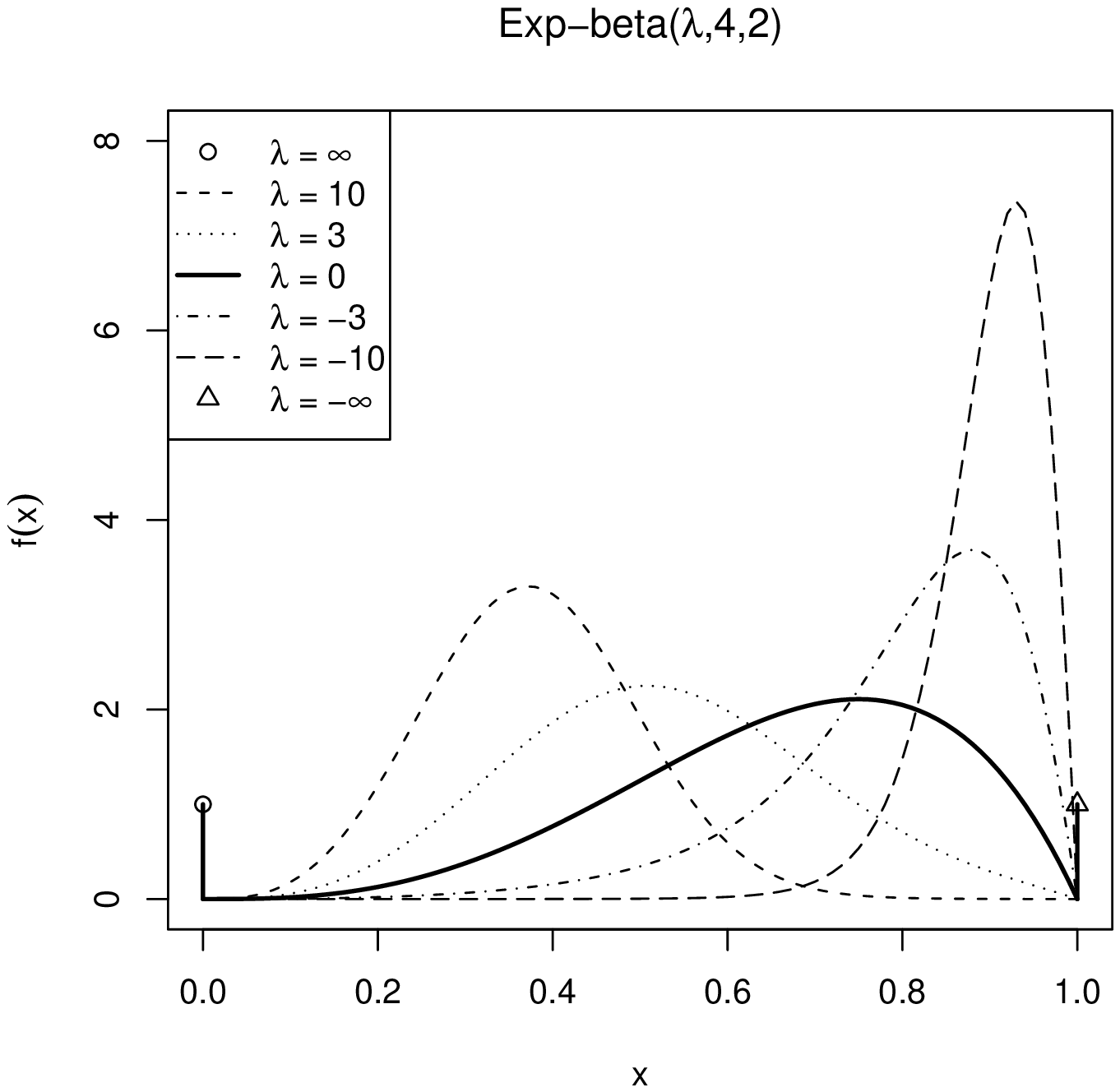}\includegraphics[width=0.45\textwidth]{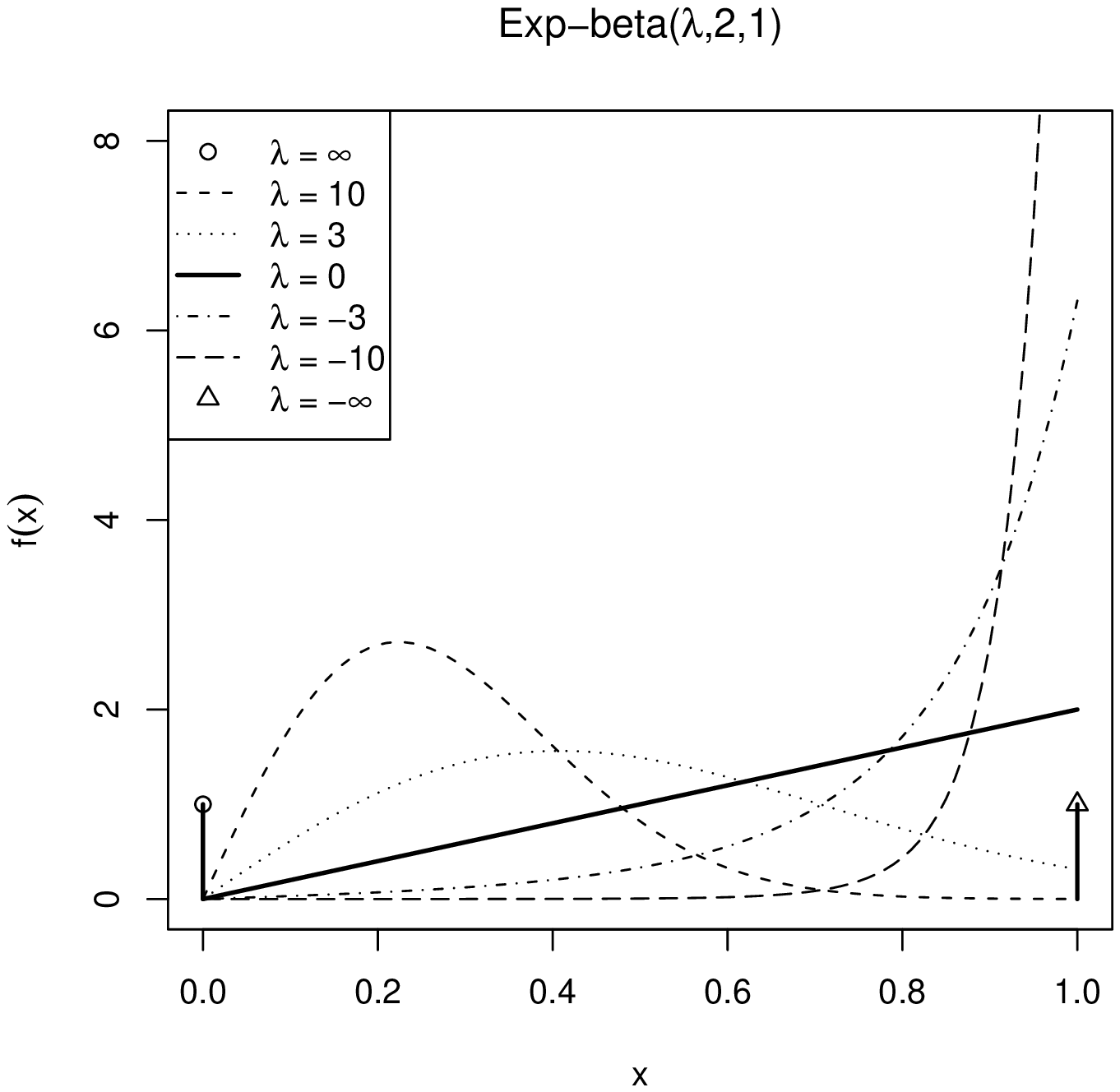}
	\includegraphics[width=0.45\textwidth]{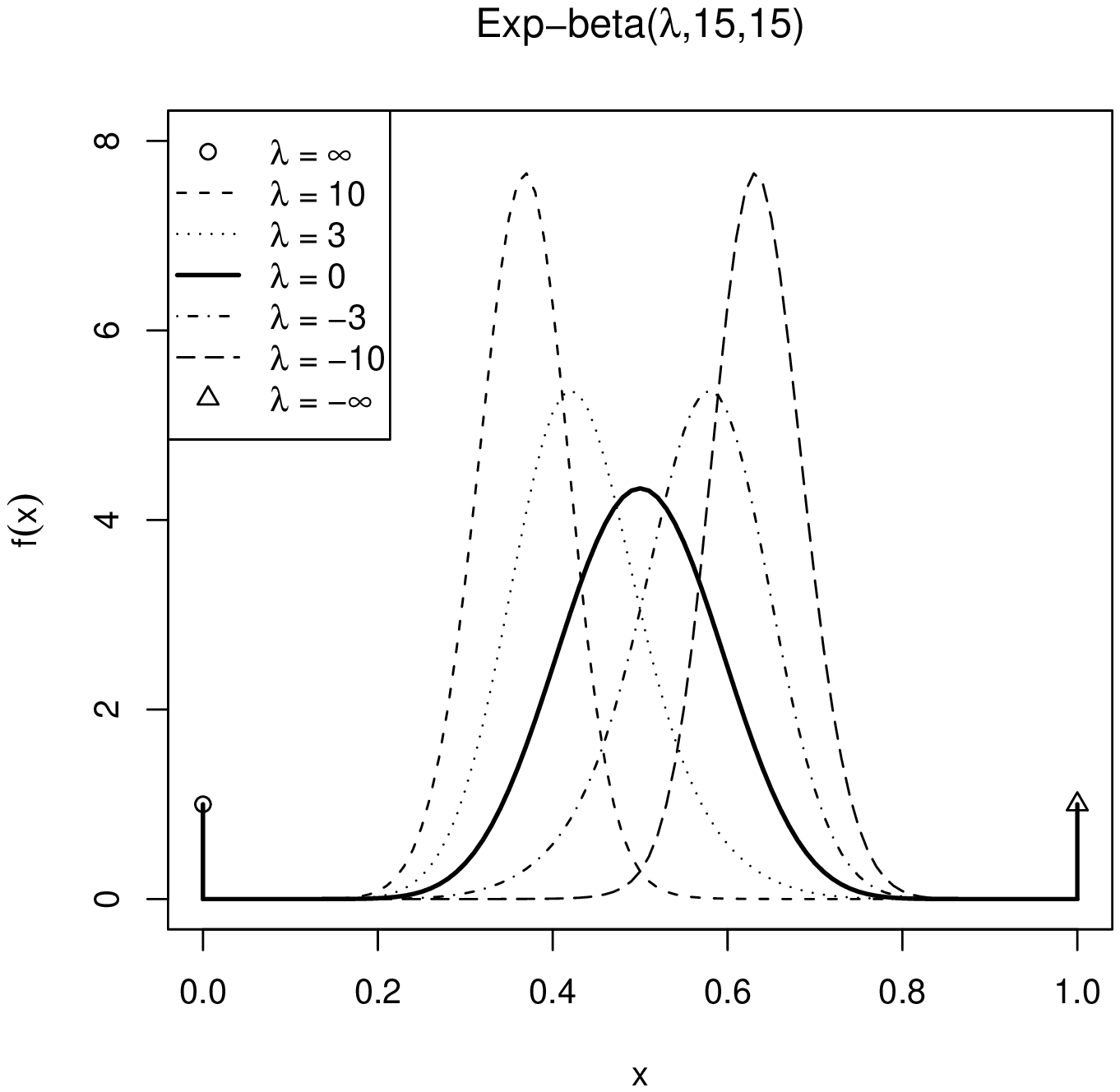}
	\caption{Graphics of the pdf of the exp-beta distribution for some values of the parameters.}
	\label{pdf2}
\end{figure}

The pdf and cdf of the exp-beta distribution are given by
\begin{eqnarray*}
f(x)=\frac{\lambda}{B(a,b)(1-e^{-\lambda})}x^{a-1}(1-x)^{b-1}e^{-\lambda I_x(a,b)}, \quad x\in (0,1)
\end{eqnarray*}
and
\begin{eqnarray*}
F(x)=\frac{1-e^{-\lambda I_x(a,b)}}{1-e^{-\lambda}}, \quad x\in (0,1),
\end{eqnarray*}
respectively.\\
Figure \ref{pdf2} shows the plots of the pdf of the exp-beta distribution for some values of $a$ and $b$, and for $\lambda=-\infty,-10,-3,0,3,10,\infty$. Observe that for the exp-beta($\lambda$,2,1) distribution, the density of the beta(2,1) distribution is very close to a straight line, whereas the densities of the exp-beta($\lambda$,2,1) distributions may assume various shapes, such as unimodal, and strictly increasing. 

The cdf of $Y$ can be written in the form (\ref{expcdf}). To see this, use the expansion 
$$(1-x)^{b-1}=\sum_{k=0}^\infty\frac{(-1)^k\Gamma(b)}{\Gamma(b-k)k!}x^k,$$
for $|x|<1$ and $b>0$ real non-integer. If $b>0$ is integer, the index $k$ in the above sum stops at $b-1$.  Hence, for beta distribution we obtain $a_k=(-1)^k\Gamma(a+b)/\{\Gamma(a)\Gamma(b-k)k!(a+k)\}^{-1}$ for $b>0$ real non-integer and $c=a$ in (\ref{expcdf}); if $b>0$ is integer, $a_k=0$ for $k>b-1$. Therefore, the pdf of the exp-beta distribution can be expanded in the form (\ref{usefulexpansion}).

\subsection{Order statistics and moments}
Let $X_1,\ldots,X_n$ be a random sample from exp-beta$(\lambda,a,b)$ distribution and denote the $i$th order statistic as $X_{i:n}$. The pdf of $X_{i:n}$, say $f_{i:n}$, is obtained from (\ref{pdforder}):
\begin{eqnarray*}
f_{i:n}(x)&=&\frac{\lambda(1-e^{-\lambda})^{-n}x^{a-1}(1-x)^{b-1}e^{-\lambda I_x(a,b)}}{B(i,n-i+1)B(a,b)}\{1-e^{-\lambda I_x(a,b)}\}^{i-1}\{e^{-\lambda I_x(a,b)}-e^{-\lambda}\}^{n-i},
\end{eqnarray*} 
for $x\in(0,1)$.\\

We now give an expression for the $r$th moment of the exp-beta distribution. Consider $X\sim\mbox{exp-beta}(\lambda,a,b)$. We seen that pdf of $X$ can be written in the form (\ref{usefulexpansion}). We have that $E(Y^v)=B(v+a,b)/B(a,b)$, for $v>0$. With this results, from (\ref{mom1}) and (\ref{momorder}) it follows easily that the moments of $X$ and $X_{i:n}$ are given by
\begin{eqnarray*}
E(X^r)=\frac{\lambda}{(1-e^{-\lambda})B(a,b)}\sum_{j,k=0}^\infty\frac{(-\lambda)^j}{j!}c_{j,k}B(r+k+jc+a,b)
\end{eqnarray*}
and
\begin{eqnarray*}
E(X_{i:n}^r)&=&\frac{\lambda(1-e^{-\lambda})^{-n}}{B(a,b)B(i,n-i+1)}\sum_{l,m=0}^\infty\sum_{j=0}^{i-1}\sum_{k=0}^{n-i}\frac{(-1)^{n+j-k-i}}{l!}\binom{i-1}{j}\binom{n-i}{k}e^{-\lambda(n-k-i)}\nonumber\\
&&\times\{-\lambda(j+k+1)\}^lc_{l,m}B(r+m+lc+a,b),
\end{eqnarray*}  
respectively. More once, we used the expansions of the Subsection (\ref{expansions}) and see its importance.

\begin{figure}[h!]
	\centering
	\includegraphics[width=0.45\textwidth]{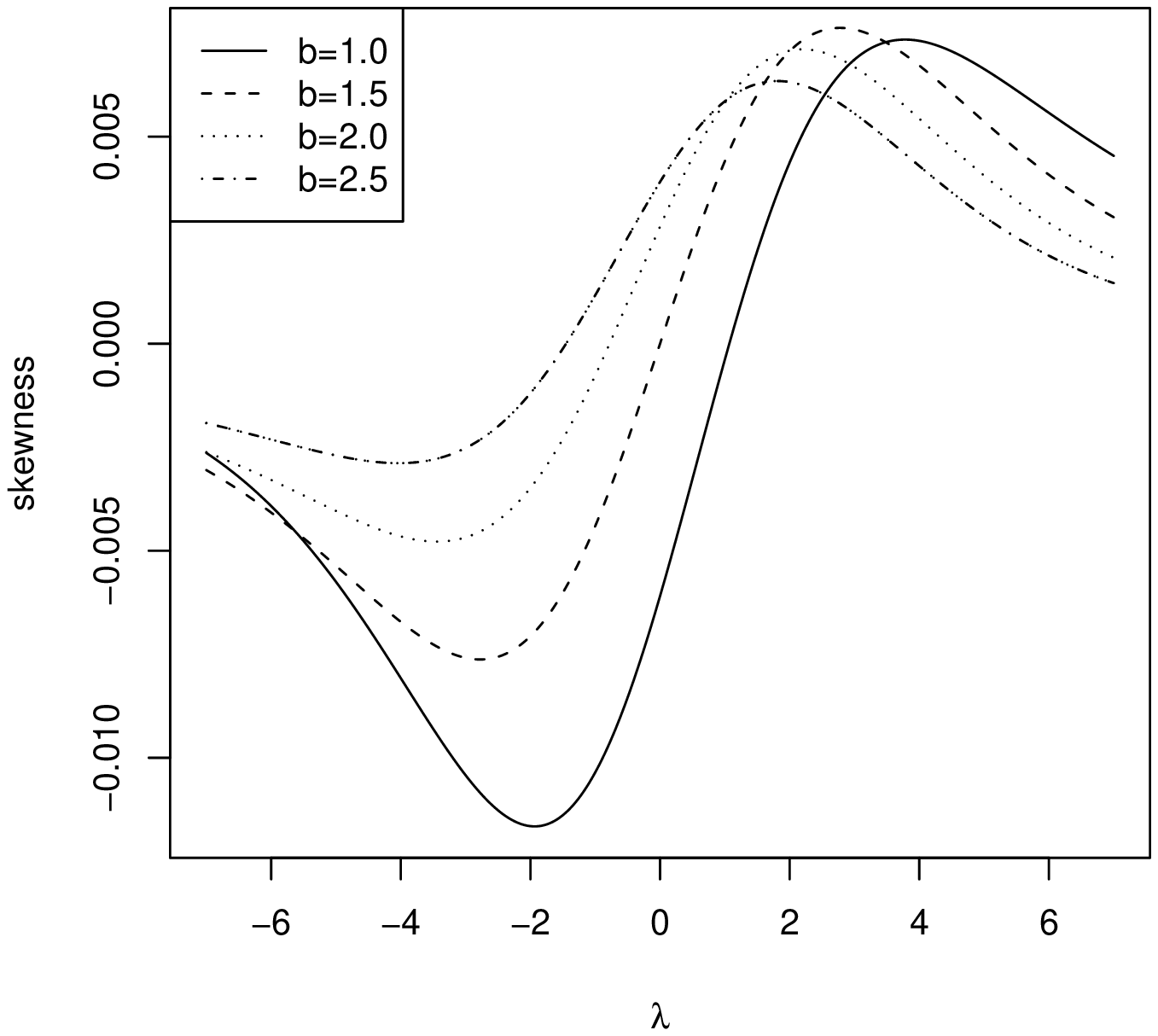}\includegraphics[width=0.45\textwidth]{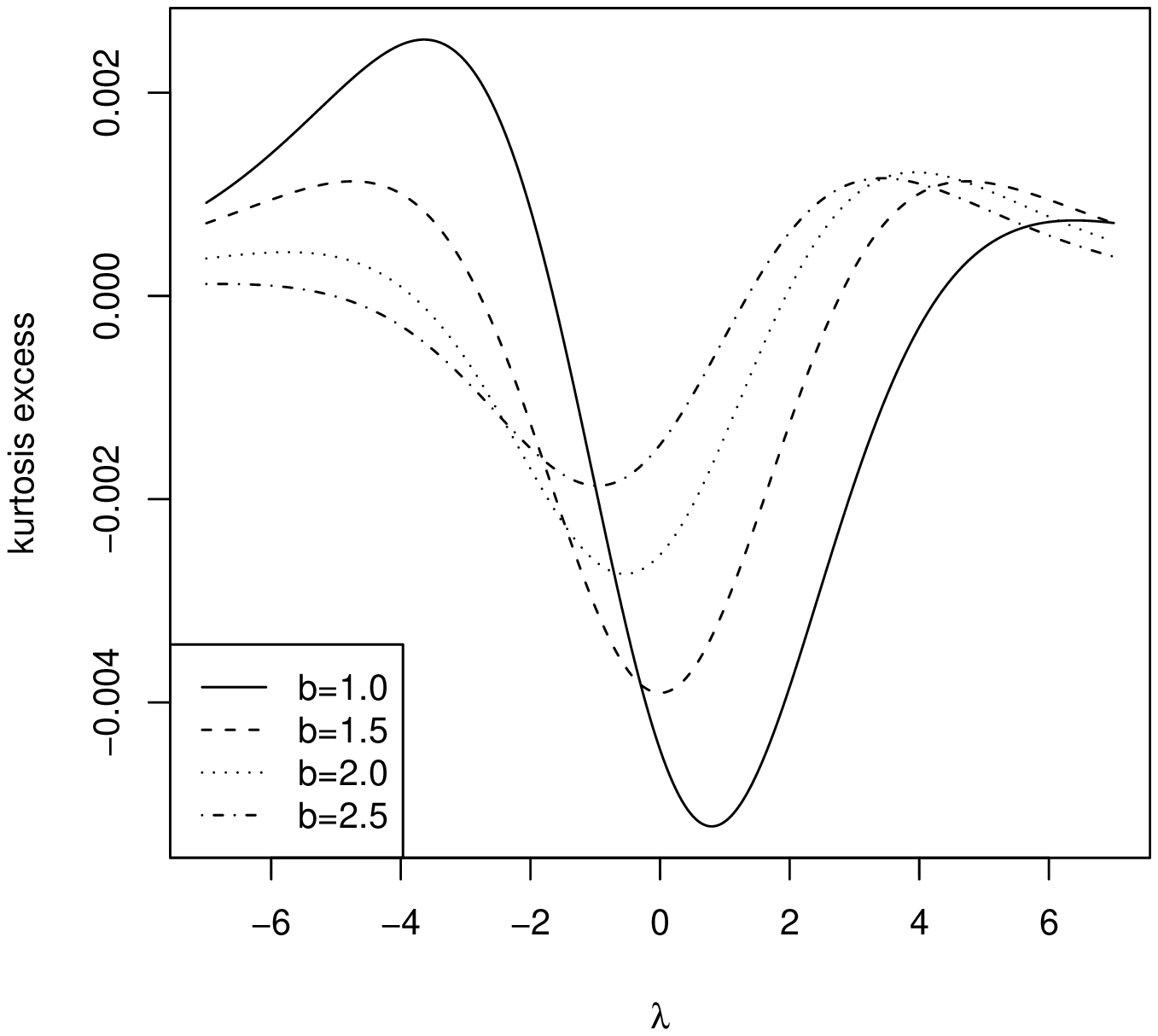}
			\caption{Skewness and kurtosis of the exp-beta distribution for some values of the parameters.}
			\label{skewkur2}
\end{figure}

Figure \ref{skewkur2} shows skewness and kurtosis of the exp-beta distribution, obtained from application of the formula of the moments above, for $a=2$ and some values of $b$ as function of $\lambda$. 

\subsection{Score function and information matrix}

Let $X$ random variable with $\mbox{exp-beta}(\lambda,a,b)$ distribution and $\theta=(\lambda,\beta,\alpha)^T$ be the parameter vector, with $\lambda\neq0$.  The log-density $\ell=\ell(\theta)$ for the random variable $X$ with observed value $x$ is given by
\begin{eqnarray*}
\ell= \log\lambda-\log B(a,b)-\log(1-e^{-\lambda})+(a-1)\log x +(b-1)\log(1-x)-\lambda I_x(a,b),
\end{eqnarray*}
for $x\in(0,1)$.\\

The score function is given by
\begin{eqnarray*}
\frac{\partial l}{\partial \lambda} &=& \frac{1}{\lambda}-\frac{1}{e^\lambda-1}-I_x(a,b), \\
\frac{\partial l}{\partial a} &=& -\Psi(a)+\Psi(a+b)+\log x-\lambda\frac{\partial I_x(a,b)}{\partial a}, \\
\frac{\partial l}{\partial b} &=& -\Psi(b)+\Psi(a+b)+\log(1-x)-\lambda\frac{\partial I_x(a,b)}{\partial b},
\end{eqnarray*}
where $\Psi(y)=d\log\Gamma(y)/dy$.\\ 

Under the usual regularity conditions, the expected value of the score function vanishes. Hence, we obtain
$$E\{I_X(a,b)\}=\frac{1}{\lambda}-\frac{1}{e^\lambda-1},$$
$$E\left\{\frac{\partial I_X(a,b)}{\partial a}\right\}=\lambda^{-1}\{\Psi(a+b)-\Psi(a)+E(\log X)\}$$
and 
$$E\left\{\frac{\partial I_X(a,b)}{\partial b}\right\}=\lambda^{-1}[\Psi(a+b)-\Psi(b)+E\{\log(1-X)\}].$$

The Fisher's information matrix $K=K((\lambda,a,b)^T)$ is
$$K = \left(\begin{array}{ccc}
\kappa_{\lambda,\lambda}&\kappa_{\lambda,a}&\kappa_{\lambda,b}\\
\kappa_{\lambda,a}&\kappa_{a,a}&\kappa_{a,b}\\
\kappa_{\lambda,b}&\kappa_{a,b}&\kappa_{b,b}\\
\end{array}\right),$$
whose elements are
$$\kappa_{\lambda,\lambda} = {\lambda }^{-2}-\frac{e^{\lambda }}{{\left( -1 + e^{\lambda } \right) }^2},\quad\kappa_{\lambda,a} =\lambda^{-1}\{\Psi(a+b)-\Psi(a)+E(\log X)\},$$
$$\kappa_{\lambda,b} = \lambda^{-1}[\Psi(a+b)-\Psi(b)+E\{\log(1-X)\}],$$
$$\kappa_{a,a} = \Psi'(a)-\Psi'(a+b)+\lambda E\left\{\frac{\partial^2 I_X(a,b)}{\partial a^2} \right\},$$
$$\kappa_{a,b} = \Psi'(b)-\Psi'(a+b)+\lambda E\left\{\frac{\partial^2 I_X(a,b)}{\partial b^2} \right\}$$
and
$$\kappa_{b,b} = -\Psi'(a+b)+\lambda E\left\{\frac{\partial^2 I_X(a,b)}{\partial a\partial b}\right\}.$$
These elements of the information matrix depend on some expectations that can be easily obtained through numerical integration. 

\section{Application}\label{sec5}
Our aim in this Section is to motivate the use of the class exp-G of distributions by showing a successful application to one real data set. We here will fit exp-Weibull distribution to the data set given by Birnbaum and Saunders (1969) on the fatigue life of 6061-T6 aluminium coupons cut parallel to the direction of rolling and oscillated at 18 cycles per second. The data set consists of 101 observations with maximum stress per cycle 31,000 psi. The data are: 70, 90, 96, 97, 99, 100, 103, 104, 104, 105, 107, 108, 108, 108, 109, 109, 112, 112, 113, 114, 114, 114, 116, 119, 120, 120, 120, 121, 121, 123, 124, 124, 124, 124, 124, 128, 128, 129, 129, 130, 130, 130, 131, 131, 131, 131, 131, 132, 132, 132, 133, 134, 134, 134, 134,
136, 136, 137, 138, 138, 138, 139, 139, 141, 141, 142, 142, 142, 142, 142, 142, 144, 144, 145, 146, 148, 148, 149, 151, 151,
152, 155, 156, 157, 157, 157, 157, 158, 159, 162, 163, 163, 164, 166, 166, 168, 170, 174, 201, 212.\\

The MLEs and the maximized log-likelihood determined by fitting the exp-Weibull and Weibull distributions are
\begin{eqnarray*}
\widehat{\beta}=55.670932, \quad \widehat{\lambda}=-41.645738,\quad \widehat{\alpha}=1.642486,\quad \widehat\ell_{exp-Weibull}=-454.3272
\end{eqnarray*}
and
\begin{eqnarray*}
\widehat{\beta}=143.3150, \quad \widehat{\alpha}=5.9790,\quad \widehat\ell_{Weibull}=-459.0999,
\end{eqnarray*}
respectively.

\begin{figure}[h!]
	\centering
		\includegraphics[width=0.50\textwidth]{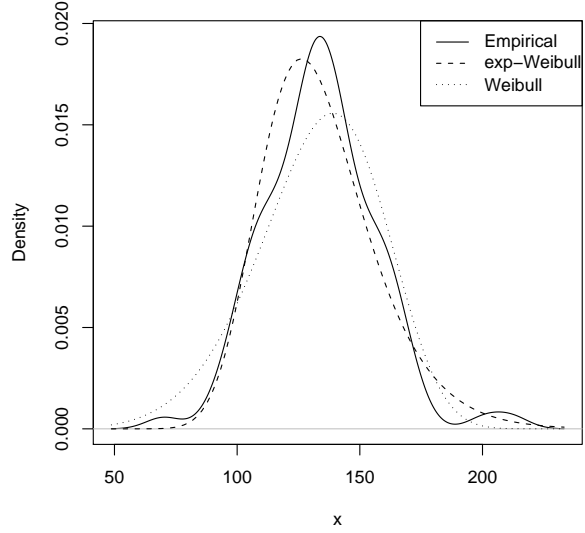}
	\caption{Empirical density and fitted exp-Weibull and Weibull densities for the Birnbaum and Saunders's (1969) data set. }
	\label{figapp}
\end{figure}

We test the null hypothesis $H_0: \mbox{Weibull model}$ against the alternative hypothesis $H_1: \mbox{exp-Weibull model}$. The LR statistic is 9.5453
and $\mbox{p-value}=2\times10^{-3}$. Hence, for any usual significance level, we reject null model (Weibull) in favour of the alternative exp-Weibull model. In Figure \ref{figapp} are displayed the empirical density, fitted exp-Weibull and Weibull densities. Hence, we see that the exp-Weibull distribution yields a better fit than Weibull distribution.

\section{Conclusion}\label{sec6}
We defined a family of distributions that provides a rather general and flexible framework for statistical analysis. It also provides a rather flexible mechanism for fitting a wide spectrum of real world data sets.

 Several properties of this class of distributions was obtained, such as Kullback-Leibler divergence between $G$ and exp-$G$ distributions,  characterization based on Shannon entropy, moments, order statistics, estimation of the parameters and inference.

With this, we moved to two special distributions, the exp-Weibull and exp-beta distributions, which were studied with some details. The article were motivated by a successful application to fatigue life data.

\section*{Appendix}
In this appendix we prove Propositions \ref{prop1} and \ref{prop3}, and Theorem \ref{teo1}.
\subsection*{Proof of Proposition \ref{prop1}} 
$i)$ and $ii)$ follow directly from equation \eqref{expg};\\
For $iii)$ suppose that $\lambda>0$, the calculation being analogous for $\lambda<0$. Then, 
\begin{eqnarray}
\frac{\exp\{-\lambda G(x-\epsilon)\}-\exp\{-\lambda G(x)\}}{(1-e^{-\lambda})\{G(x)-G(x-\epsilon)\}} &=& \frac{1}{(1-e^{-\lambda})\{G(x)-G(x-\epsilon)\}} \int_{G(x-\epsilon)}^{G(x)} \lambda e^{-\lambda x} dx\nonumber\\
&\leq& \frac{\lambda}{1-e^{-\lambda}} \exp\{-\lambda G(x-\epsilon)\}.\label{desig1}
\end{eqnarray}
A similar computation shows that 
$$\frac{\exp\{-\lambda G(x-\epsilon)\}-\exp\{-\lambda G(x)\}}{(1-e^{-\lambda})\{G(x)-G(x-\epsilon)\}} \geq \frac{\lambda}{1-e^{-\lambda}} \exp\{-\lambda G(x)\}.$$
Therefore, if $x$ is a continuous point of $G$, we have that
$$\frac{1}{1-e^{-\lambda}}\lim_{\substack{\epsilon\to 0\\\epsilon>0}} \frac{\exp\{-\lambda G(x-\epsilon)\}-\exp\{-\lambda G(x)\}}{G(x)-G(x-\epsilon)} = \frac{\lambda}{1-e^{-\lambda}} \exp\{-\lambda G(x)\}.$$
Nevertheless, if $x$ is a discontinuity point of $G$, we have that
$$\frac{1}{1-e^{-\lambda}}\lim_{\substack{\epsilon\to 0\\\epsilon>0}} \frac{\exp\{-\lambda G(x-\epsilon)\}-\exp\{-\lambda G(x)\}}{G(x)-G(x-\epsilon)} = \frac{\exp\{-\lambda G(x-)\}-\exp\{-\lambda G(x)\}}{(1-e^{-\lambda})\{G(x)-G(x-)\}}.$$
Suppose that $G$ is discontinuous at the points $\{x_1,\ldots\}$, and let $G_c$ be the part of $G$ with continuity points (the sum of the continuous and singular parts of $G$), then, it is easy to observe that
$$F_\lambda^G(x) = F_\lambda(G_c(x)) + \sum_{i=1}^\infty 1_{\{x_i\leq x\}} \frac{\exp\{-\lambda G(x_{i-1})\}-\exp\{-\lambda G(x_i)\}}{1-e^{-\lambda}},$$
where $1_A(x)$ is the indicator function of the set $A$. We now, have that
\begin{eqnarray*}
\int_{(-\infty,x]} &\frac{1}{1-e^{-\lambda}}&\lim_{\substack{\epsilon\to 0\\\epsilon>0}}\frac{\exp\{-\lambda G(x-\epsilon)\}-\exp\{-\lambda G(x)\}}{G(x)-G(x-\epsilon)}dG(x)\\
&=& \int_{(-\infty,x]} \frac{\lambda}{1-e^{-\lambda}} \exp\{-\lambda G_c(x)\}dG_c(x)\\
&+& \sum_{i=1}^\infty 1_{\{x_i\leq x\}} \frac{\exp\{-\lambda G(x_{i-1})\}-\exp\{-\lambda G(x_i)\}}{1-e^{-\lambda}}\\
&=& \frac{1-\exp\{-\lambda G_c(x)\}}{1-e^{-\lambda}} + \sum_{i=1}^\infty 1_{\{x_i\leq x\}} \frac{\exp\{-\lambda G(x_{i-1})\}-\exp\{-\lambda G(x_i)\}}{1-e^{-\lambda}},
\end{eqnarray*}
which concludes the proof of $iii)$. The proof of $iv)$ is a simple application of $iii)$, and to prove $v)$, one uses the $iii)$ and the inequality in equation \eqref{desig1}, for $\lambda>0$, and a similar inequality for $\lambda<0$. \\
${}$ \hfill $\square$

\subsection*{Proof of Proposition \ref{prop3}}

Let $z(\cdot)$ be a pdf which satisfies the constraints C1 and C2. The Kullback-Leibler divergence between $z$ and $f$ is
\begin{eqnarray*}
D_{KL}(z||f)=\int_{\mathbb{R}}z\log\left(\frac{z}{f}\right)dx.
\end{eqnarray*}
With this, we follow Cover and Thomas (1991) and obtain
\begin{eqnarray*}
0\leq D_{KL}(z||f)&=&\int_{\mathbb{R}}z\log zdx-\int_{\mathbb{R}}z\log fdx\\
&=&-\mathbb{H}_{S}(z)-\int_{\mathbb{R}}z\log fdx.
\end{eqnarray*}
With the definition of $f$ and based on the constraints C1 and C2, it is easy to see that 
\begin{eqnarray*}
\int_{\mathbb{R}}z\log fdx&=&-1+\frac{\lambda }{e^\lambda-1}+\log\left(\frac{\lambda}{1-e^{-\lambda}}\right)+E\{\log g(G^{-1}(U;\theta))\}\\
&=&\int_{\mathbb{R}}f\log fdx=-\mathbb{H}_{S}(f),
\end{eqnarray*}
where $U$ is defined as before. With this, we have 
$$\mathbb{H}_{S}(z)\leq\mathbb{H}_{S}(f),$$
with equality if and only if $z(x)=f(x)$ Lebesgue - almost everywhere, thus proving the uniqueness.
${}$\hfill$\square$

\subsection*{Proof of Theorem \ref{teo1}}
(i) It is well-known from real analysis that if $f_n$ is a sequence of bounded and right-continuous functions that converge in all point for a continuous function, then this convergence is uniform. Therefore, since $F_\lambda$ satisfies the above conditions and converges for $G$, which is a continuous function, the proof of (i) is completed.\\

\noindent(ii)-(iv) The proofs are easily checked.\\

\noindent(v) For $(\lambda_0,\theta_0)\in\Theta$, with $\lambda_0\in\mathbb{R}\setminus\{0\}$, the result follows of the Proposition \ref{prop2} and the Theorem 6.5.1 from Lehmann and Casella (2003). For $\lambda_0=0$, we have to use the results in (ii)-(iv) to adapt the proof given in Lehmann  and Casella (2003) to our case. We begin by showing that
$$\sqrt{n}\hat\lambda\stackrel{d}{\rightarrow} N(0,12),$$
when $n\rightarrow\infty$. The following lemma will be useful in this proof.\\

\begin{lemma} Let $f:[a,b)\rightarrow\mathbb{R}$ ($b$ can be $\infty$). If $f$ admits $n$ derivatives to right around the point $a$, then 
$$f(x)=f(a)+f^{(1)}(a)(x-a)+\ldots+\frac{f^{(n)}(a)}{n!}(\tilde{x}-a)^n,$$ 
where $f^{(k)}$ represents the $k$th derivative of $f$ to right and $\tilde{x}\in(a,x)$.
\end{lemma}

The proof the above lemma is similar to usual proof, but replacing the derivative by the derivative from the right. It is clear that the same result holds for derivative from the left.

Applying the Lemma to log-likelihood with respect to pdf $f(x;\lambda)=\lambda e^{-\lambda x}/(1-e^{-\lambda})$, i.e. $G(x;\theta)=x$, if follows that 
\begin{eqnarray*}
\ell'(\hat\lambda)&=&\ell'(0)+\hat\lambda \ell''(0)+\frac{1}{2}\hat\lambda^2\ell'''(\tilde\lambda)\\
&=&\frac{n}{2}-\sum_{j=1}^n x_j-\frac{n}{12}\hat\lambda+\frac{1}{2}\hat\lambda^2\ell'''(\tilde\lambda),
\end{eqnarray*}
where $\ell'''(\tilde\lambda)=-\frac{1}{6}\tilde\lambda^{-3}-d^3(e^{\tilde\lambda}-1)^{-1}/d\tilde\lambda$.

By supposition, $\ell'(\hat\lambda)=0$, thus
$$\sqrt{n}\hat\lambda=n^{-1/2}\frac{n/2-\sum_{j=1}^nx_j}{12^{-1}-(2n)^{-1}\hat\lambda\ell'''(\tilde\lambda)}.$$
As $n^{-1}\hat\lambda\ell'''(\tilde\lambda)\rightarrow0$ in probability and $n^{-1/2}(n/2-\sum_{j=1}^nx_j)\stackrel{d}{\rightarrow} N(0,1)$, when $n\rightarrow\infty$.\\

The rest of the proof is analogous to the one given in Lehmann, where one may use the results in (ii)-(iv) to ensure that all the arguments holds true. 

\noindent(vi) It follows from asymptotic normality of $\hat\Theta$, see Lehmann and Romano (2008) for more details. \\
${}$\hfill$\square$

\end{document}